\documentclass[article]{IEEEtran}
\usepackage[cmex10]{amsmath}
\usepackage[short]{optidef}
\usepackage{color}
\usepackage{commath}
\usepackage{algorithm}
\usepackage{algorithmic}
\usepackage{textcomp}
\usepackage{gensymb}
\usepackage{multirow}
\usepackage{multicol}
\usepackage{mathtools}
\ifCLASSINFOpdf
   \usepackage[pdftex]{graphicx}
\graphicspath{ {figures/} }
\else
\fi
\usepackage{graphicx}
\usepackage[font=footnotesize,caption=false,subrefformat=parens,labelformat=parens]{subfig}
\renewcommand{\figurename}{Fig.}
\usepackage{cite}
\usepackage{amssymb}
\usepackage{amsthm}
\usepackage{url}
\usepackage{pifont}
\newcommand{\cmark}{\ding{51}}%
\newcommand{\xmark}{\ding{55}}%

\usepackage{siunitx}

\DeclarePairedDelimiter\floor{\lfloor}{\rfloor}  

\usepackage{eucal}
\usepackage{enumitem}

\newcommand\blfootnote[1]{%
		\begingroup
		\renewcommand\thefootnote{}\footnote{#1}%
		\addtocounter{footnote}{-1}%
		\endgroup
}

\title{3D Trajectory Optimization in UAV-Assisted Cellular Networks Considering Antenna Radiation Pattern and Backhaul Constraint}

\author{Md Moin Uddin Chowdhury,
Sung Joon Maeng, Eyuphan Bulut, and \.{I}smail G\"{u}ven\c{c} 
\thanks{M.M.U. Chowdhury, S.J. Maeng, and \.{I}. G\"{u}ven\c{c} are with the Department of Electrical and Computer Engineering, North Carolina State University, Raleigh, NC (e-mail:~\{mchowdh,smaeng,iguvenc\}@ncsu.edu).}
\thanks{E. Bulut is with the the Department of Computer Science, Virginia Commonwealth University, Richmond, VA, USA (e-mail:~ebulut@vcu.edu).}}

\begin{document}
\pdfoutput=1
\maketitle
\begin{abstract}

This paper explores the effects of three-dimensional (3D) antenna radiation pattern and backhaul constraint on optimal 3D path planning problem of an unmanned aerial vehicle (UAV), in interference prevalent downlink cellular networks. We consider a cellular-connected UAV that is tasked to travel between two locations within a fixed time and it can be used to improve the cellular connectivity of ground users by acting as a relay. Since the antenna gain of a cellular base station changes significantly with the UAV altitude, the UAV can improve the signal quality in its backhaul link by changing its height over the course of its mission. This problem is non-convex and thus, we explore the dynamic programming technique to solve it. We show that the 3D optimal paths can introduce significant network performance gain over the trajectories with fixed UAV heights. 
\end{abstract}

\section{Introduction}

\blfootnote{This research was supported by NSF under the grant CNS-1453678. Part of this work was presented at IEEE Aerospace Conference in 2019 \cite{moin2}.} 

It is estimated that the market for commercial unmanned aerial vehicles (UAVs) will reach around 13 billion U.S. dollars by the year 2025~\cite{estimate}. With the progress in the UAV manufacturing technology and new regulations that will allow integration of UAVs into the allowed airspace, UAVs can be used for numerous application such as package delivery, surveillance, and agriculture. In addition to these use cases, UAVs can also serve as wireless base stations (BSs) or relays with good line-of-sight (LOS) connectivity to both macrocell BSs (MBSs) and the ground user equipment (UEs). Using UAVs as aerial network nodes can enable flexible and rapid deployments for providing on-demand communications in cellular hot-spot areas, and provide emergency service at disaster-affected areas. 

While dedicated UAVs as wireless BSs/relays can assist cellular service providers to achieve better network performance, the vast number of other UAVs, such as those that are tasked for parcel delivery, can concurrently provide wireless connectivity to an underlying cellular network. They can serve as relays in a cellular network to enhance the coverage and capacity of UEs while handling their primary tasks, e.g., delivery of supplies between two locations. Since aerial package delivery will become ubiquitous in the upcoming years (e.g., due to  Amazon package delivery), use of such UAVs can reduce capital and operational expenses (CAPEX) for cellular technology, and improve energy efficiency of future wireless networks. On the other hand, the existing MBSs are equipped with antennas which are downtilted for maximizing the connectivity of the UEs. The radiation pattern of such antenna orientation varies with the height of UAV flying in the sky \cite{garca}, causing coverage problems for relay/backhaul links of UAVs\cite{backhaul}. 
In order to get better antenna gain in the backhaul link, UAVs may need to change their heights during the predefined mission for providing better network coverage to UEs. Since UAVs are battery-limited devices, to exploit the full potential of the UAVs as flying relay BSs, optimal 3D path planning is of critical importance.  
\begin{figure}[t]
\centering
\vspace{-3mm}
\includegraphics[width=1\linewidth]{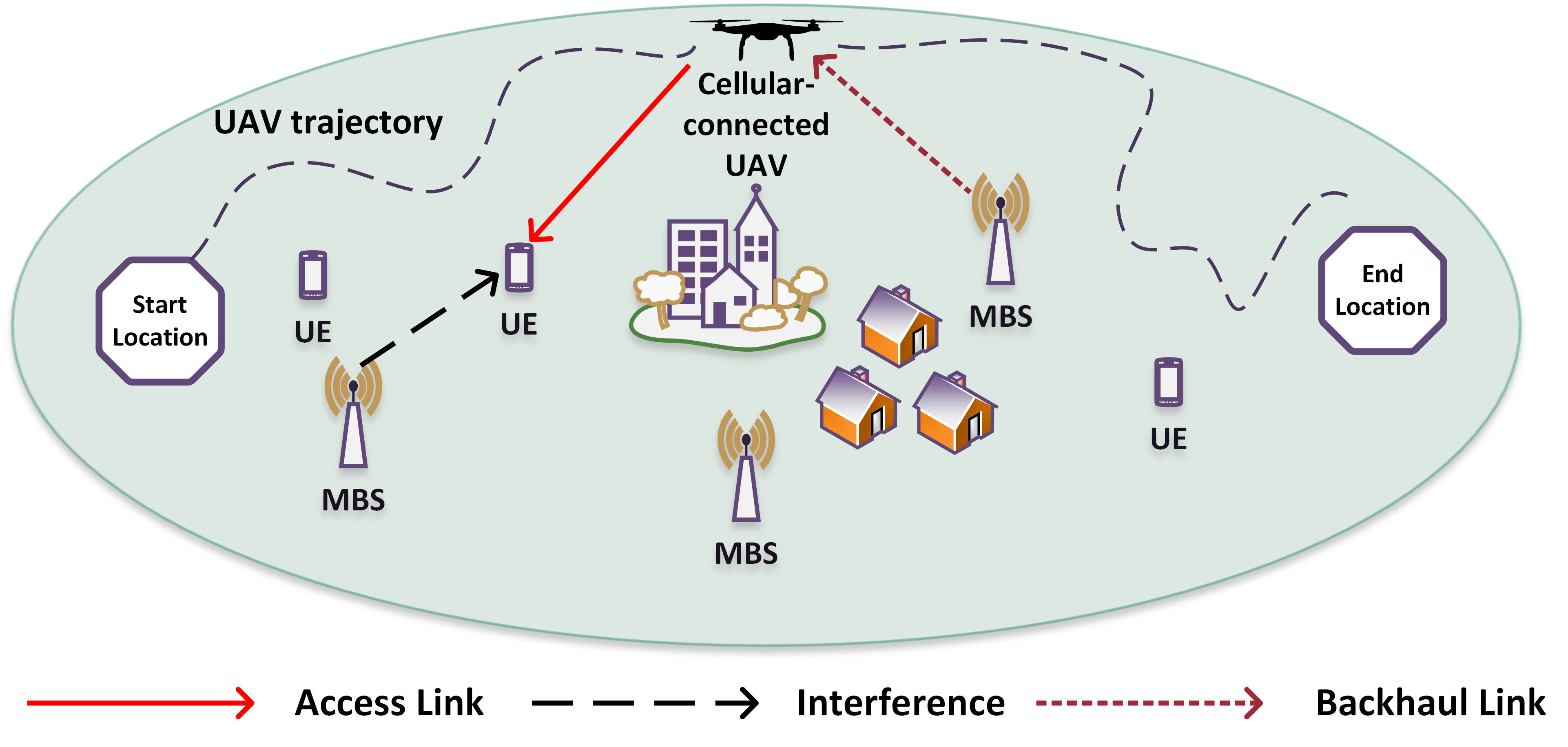}\\
\caption{{A typical scenario where a UAV is flying from start location towards end location to fulfill mission requirement. While traversing, it will provide downlink coverage while maintaining backhaul connection with the MBSs. The optimal trajectory of the UAV will be effected by the quality of the backhaul link and 3D antenna radiation of the MBS antennas.}}
\label{network}
\end{figure}

While integrating UAVs in wireless networks to enhance network capability has recently received extensive interest \cite{dp,rajeev,rui1,rui2,multiuav}, the number of works considering realistic interference prevalent cellular network scenario is surprisingly small. This motivated us to study the optimal path planning for a UAV acting as a wireless relay for improving the terrestrial downlink cellular network performance while considering realistic 3D antenna radiation pattern in the relay/backhaul link, specified by 3GPP \cite{ant3gpp}. More specifically, we consider a UAV with a primary task, such as the delivery of supplies among two locations within a given time budget. During its mission, the UAV can be used to offload cellular users from the MBSs in order to improve the overall network capacity. A high-level overview of such a scenario is depicted in Fig.~\ref{network}.

In contrast to our previous works in \cite{moin2} and \cite{moin}, which consider only a fixed height for UAVs, we consider the 3GPP antenna radiation model which is dependent on the elevation and azimuth angles between an MBS and a UE. Hence, we consider a 3D trajectory optimization problem where the UAV can change both the height and the trajectory along the horizontal plane in order to enhance its received signal quality of the backhaul link. Our contributions in this work can be summarized as follows:

\begin{itemize}
    \item We formulate a time-constrained 3D trajectory optimization problem by considering both backhaul constraint\footnote{In this work, we did not consider maintaining a minimum capacity in the backhaul link for simplicity. We leave this problem for future.} and 3D antenna radiation pattern of MBSs in an in-band downlink cellular network, where the locations of the ground users (UE) and MBSs are distributed uniformly across the network. 
    
    \item We provide a detailed review of realistic 3D antenna pattern and relevant path loss models that have critical impact on the trajectory optimization problem.
    
    \item Since the trajectory optimization problem is non-convex, it is difficult to solve in general. Therefore, we use dynamic programming (DP) to obtain the optimal approximate paths. We divide time, possible actions, and 3D space into discrete steps. At a certain time step, the UAV can choose any action from a discrete action set which will move the UAV to a new location. This kind of approximation will help us to find the optimal trajectories efficiently.
    
    \item We run extensive Matlab simulations to gain insights on the effects of both backhaul and antenna radiation pattern on the network performance associated with the optimal 3D trajectories. We compare network performances of 3D path planning with those of 2D optimal trajectory and show that the 3D trajectory provides higher network performance gain than its 2D counterpart. The ability of the UAV to change its altitude helps it to maintain better connectivity in the backhaul link which results in better network performance.
    
\end{itemize}

The rest of the paper is organized as follows. In Section~\ref{literature}, we provide a literature review related to UAV assisted wireless networks.~Section~\ref{sys_mod} illustrates the system model. We discuss some factors which can influence UAV trajectories in cellular networks in Section~\ref{sec:factors}. The optimal path planning problem is discussed in Section~\ref{sec:prob}. In Section~\ref{sec:dp}, we discuss our DP based approach for solving 3D trajectory optimization problem. Numerical results and simulation parameters are presented in Section~\ref{sec:simulation}, and Section~\ref{sec:conclusion} concludes our paper.
\begin{table*}[t]
\centering
\caption{{Literature review for UAV trajectory optimization.}} 
\scalebox{0.98}{
\begin{tabular}{p{0.6cm} p{2cm} p{6cm} p{2cm} p{1cm} p{1.5cm} p{1cm} } \hline
{Ref.} & {Technique} & {Goal} & {Cellular-connected UAV} & {3D trajectory planning} & {Backhaul constraints} & {3D antenna radiation} \\ \hline
\cite{Bulut} & DP & Mission completion time minimization with disconnectivity duration constraint & \cmark & \xmark & \xmark & \xmark \\ \hline

\cite{rajeev} & DP & Maximize sum-rate of users& \xmark & \xmark & \xmark & \xmark\\ \hline
\cite{gesbert} & DP & Study the trade-off between sum-rate and recharging battery & \xmark & \xmark & \xmark & \xmark\\ \hline
\cite{moin} & DP & Study network performance of different scheduling criterion & \cmark & \xmark & \xmark & \xmark\\ \hline
\cite{moin2} & DP & Study effects of different path loss models, backhaul constraint and antenna radiation & \cmark & \xmark & \cmark & \xmark\\ \hline
\cite{adhoc} & Minimum weighted Sum & Distributed path planning
with time constraints using multiple UAV & \xmark & \xmark & \xmark & \xmark\\ \hline
\cite{rui1}& Convex optimization & Maximize the minimum throughput in multi-UAV scenario & \xmark & \xmark & \xmark & \xmark\\ \hline
\cite{rui2}&  Convex optimization & Minimize hovering time for multicasting & \xmark & \xmark & \xmark & \xmark\\ \hline
\cite{saad}& Deep reinforcement learning & Mitigate interference & \cmark & \xmark & \xmark & \xmark\\ \hline
\cite{jiang}&  Line-search & Maximize ergodic sum-rate of an uplink wireless network & \xmark & \xmark & \xmark & \xmark\\ \hline
\cite{merwaday} & Genetic algorithm & Optimize UAV locations and interference management parameters after disaster & \cmark & \xmark & \xmark & \xmark\\ \hline
\cite{ali} & Weighted cheeger constant & Mitigate interference from
the primary network in a cognitive radio scenario & \xmark & \cmark & \xmark & \xmark\\ \hline
\cite{murat} & Mixed integer linear programming & Optimize the transmit power and trajectories of the relaying UAVs & \cmark & \cmark & \xmark & \xmark\\ \hline
\cite{wpt} & Convex optimization &  Maximize the energy transferred during a finite charging period & \xmark & \xmark & \xmark & \xmark\\ \hline
\cite{EE} & Convex optimization & Maximize the energy efficiency (EE) & \cmark & \cmark & \xmark & \xmark\\ \hline
\cite{quant} & Quantization theory & Optimal deployment and movement of multiple UAVs & \xmark & \xmark & \xmark & \xmark\\ \hline
\cite{secracy} & Convex optimization & Maximize the average worst-case secrecy & \xmark & \xmark & \xmark & \xmark\\ \hline
\cite{dual} & Penalty
concave-convex procedure & Maximize the minimum average secrecy rate & \xmark & \xmark & \xmark & \xmark\\ \hline
\cite{power} & Convex optimization & Maximize the end-to-end throughput of an amplify and forward relay & \xmark & \xmark & \xmark & \xmark\\ \hline
\cite{multiuav} & Convex optimization & Maximize the total sum-rate of users by minimizing interference among multiple UAVs & \xmark & \cmark & \xmark & \xmark\\ \hline
\cite{reinf} & Reinforcement learning & Maximize the sum-rate of users in a multi-UAV scenario with trajectory planning and power control & \cmark & \cmark & \xmark & \xmark\\ \hline
\cite{solar} & Monotonic optimization & Maximize the total rate of by jointly optimizing trajectory and resource allocation of a solar powered UAV & \xmark & \cmark & \xmark & \xmark\\ \hline
\cite{IAB:FIU} & Fixed point \& particle swarm optimization  & Maximize sum-rate by managing interference between the access links and the backhaul links & \cmark & \xmark & \cmark & \xmark\\ \hline
\cite{farshad} &  Cramer-Rao lower bound & Tracking a moving RF source  & \xmark & \xmark & \xmark & \xmark\\ \hline
\cite{moin3} & Reinforcement learning & Indoor UAV navigation & \xmark & \xmark & \xmark & \xmark\\ \hline
This work & Dynamic programming & Maximize the total rate of users in downlink & \cmark & \cmark & \cmark & \cmark\\ \hline
\end{tabular}}
\label{lit_review}
\end{table*}

\section{Related Works}
\label{literature}

Existing cellular networks are optimized for ground users with the antennas of base stations being downtilted to optimize the ground coverage and to reduce the inter-cell interference~\cite{eric}. In such a case, UAVs flying in the sky may be served by the sidelobes of base station antennas which provide smaller antenna gains\cite{garca}. Moreover, for reliable and safe operations of autonomous UAVs in beyond visible line-of-sights (BVLoS) scenarios, maintaining a backhaul with the core network is a must~\cite{backhaul}.

UAV-enabled wireless networks have already been studied widely for maximizing the sum-rate of the networks by exploiting mobility and ease of deployment of the UAVs along with the favorable UAV-ground communication \cite{gesbert,rui1,jiang,solar,reinf,multiuav}. In particular, by controlling the mobility of the UAVs, the channel between the UAV and the intended users can be improved extensively. In~\cite{rajeev}, authors maximize the weighted sum-rate of an uplink wireless network by means of optimal UAV path planning. The optimal trajectory was calculated using DP and analysis of the optimal trajectory considering an infinite hovering time is provided. Optimal path planning problem of multiple UAVs is rigorously investigated in \cite{rui1}, where the minimum throughput of the users is maximized by jointly optimizing user scheduling and transmit power of the UAVs. In \cite{jiang}, authors study the heading angle optimization using the line search method in a predefined UAV trajectory for maximizing the uplink sum-rate. Considering multi-UAV scenarios, the sum-rate of users is maximized in \cite{multiuav} and in \cite{reinf} by minimizing the interference between the UAVs and by exploiting power control technique, respectively. In a recent study, authors jointly optimize the 3D trajectory and resource allocation to get the maximum sum-rate in a downlink solar-powered UAV scenario \cite{solar}. They explore monotonic optimization for solving the problem and consider an out-of-band free space optical (FSO) link for backhaul access. However, none of these studies consider underlying cellular networks and backhaul constraint. 

UAVs are also considered as a relay component for increasing network performance. For instance, maximizing the end-to-end throughput of a mobile relay system by optimizing the source/relay power allocation and the UAV’s trajectory, is discussed in \cite{power}. Using sequential convex optimization technique, authors study the throughput maximization problem in mobile relaying systems by optimizing the transmit power in \cite{rui6}. They also consider the information-causality constraint at the relay. However, the effect of antenna radiation pattern on the relay link is not considered in either studies. On the other hand, authors in \cite{garca} investigate the feasibility of massive MIMO (mMIMO) for consistent UAV command and control (C2) support and throughput performance by means of extensive 3GPP compliant simulations. Their study shows promising result for supporting UAVs as relays in mMIMO based cellular networks. 
\looseness=-1

Trajectory optimization of UAVs is also investigated in various other contexts such as mitigating interference in cognitive radio setup \cite{ali} and in ground network \cite{saad}, minimizing hovering time for multicasting \cite{rui2}, maximizing energy efficiency \cite{EE}, tracking an RF source \cite{farshad}, indoor search and rescue operation\cite{moin3}, among others. A joint trajectory and transmit power design algorithm is proposed in \cite{secracy} to maximize the average worst-case secrecy, while in \cite{dual}, authors propose a UAV enabled secure communication system, where one UAV serves multiple users while the other UAV flies to jam the eavesdroppers on the ground. In~\cite{gesbert}, authors consider landing spots for UAV  battery recharging to study the trade-off between throughput and battery power using DP. Authors in \cite{moin} exploit DP to find optimal trajectories for different scheduling criterion in realistic interference prevalent downlink cellular networks. In \cite{Bulut}, authors aim to find the optimal UAV trajectory with the shortest duration by using DP \cite{dp}, while allowing short-term outage due to the presence of coverage holes in the networks. Apart from these, static UAV placement optimization as quasi-stationary BS is studied in \cite{merwaday,saad2,quant} for providing better connectivity to the users from a fixed altitude. 

In a recent work, authors explore the interference management problem in a UAV-assisted network by considering the mutual interference between the access links and the backhaul links\cite{IAB:FIU}. Using fixed-point method and bio-inspired heuristic technique, they maximize the network sum-rate by jointly optimizing power and 3D UAV placement. However, these works have not considered 3D trajectory optimization problem while taking the 3D antenna radiation pattern and backhaul constraint into account simultaneously. We summarize and compare some of the papers in Table~\ref{lit_review} for the convenience of readers.

\section{System Model}
\label{sys_mod}
\subsection{Network and UAV Mobility Model}
In this paper, we consider a UAV that is flying in 3D Cartesian coordinate with a maximum speed of $V_{\mathrm{max}}$ in a suburban environment as shown in Fig.~\ref{network}. The UAV has a mission to complete within a certain time-frame. To complete the mission, it has to fly from a start location, $L_{\rm s}$ to a final destination point $L_{\rm f}$ within a fixed time $T$ over an area of $\mathcal{A}$ $\text{km}^2$. Let us also assume that there are $M$ MBSs and $K$ static UEs with similar height $h\textsubscript{ue}$. For simplicity, we also assume that all MBSs have equal altitudes $h\textsubscript{bs}$ and transmission power $P\textsubscript{mbs}$. The set of the UEs can be denoted as $\mathcal{K}$ with horizontal coordinates  $\textbf{q}_k=[x_k,y_k,z_k]^T \in \mathbb{R} \textsuperscript{3x1} , {k} \in \mathcal{K}$. The MBS and the UE locations are uniformly distributed in the simulation area with densities  $\lambda_\textsubscript{mbs}$ and $\lambda_\textsubscript{ue}$ per $\text{km}^2$, respectively.

We assume that the UEs and the UAV are equipped with omnidirectional antennas and each UE associates with the strongest MBS or the UAV which acts as a relay. The MBSs are assumed to be consisting of three sectors separated by 120\degree,~while each sector is equipped with $8 \times 1$ cross-polarized antennas downtilted by 6\degree\cite{3gpp}.

 The time-varying 3D Cartesian coordinate of the UAV at time instant $t$ is denoted by $\mathbf{r(t)}=[x(t),y(t),z(t)]^T \in \mathbb{R} \textsuperscript{3x1}$ with $0\leq t\leq T$. Let us consider [$x_{\rm s},y_{\rm s},z_{\rm s}$] and [$x_{\rm f},y_{\rm f},z_{\rm f}$] to be the 3D Cartesian coordinates of $L_{\rm s}$ and $L_{\rm f}$, respectively. The minimum time required for the UAV to reach $L_{\rm f}$ from $L_{\rm s}$ with the maximum speed ${V_{\mathrm{max}}}$ is given by
\begin{equation}
  {T_{\mathrm{min}}} = \frac {\sqrt{(x_{\rm s}-x_{\rm f})^2+(y_{\rm s}-y_{\rm f})^2+(z_{\rm s}-z_{\rm f})^2})}{V_{\mathrm{max}}}. 
 \label{eqn:tmin} 
\end{equation}

The UAV's instantaneous mobility at time $t$, can be modeled as follows:
\begin{align}
\dot{x}(t)&=v(t)\sin\theta(t)\cos\phi(t),\\
\dot{y}(t)&=v(t)\sin\theta(t)\sin\phi(t),\\
\dot{z}(t)&=v(t)\cos\theta(t),
\end{align}
where $\dot{x}(t)$, $\dot{y}(t)$, and $\dot{z}(t)$ are the time derivatives of $x(t)$, $y(t)$, and $z(t)$, respectively, $v(t)$ is the velocity, $\theta(t)$ is the elevation angle with $0\leq\theta(t)\leq \pi$, and $\phi(t)$ is the azimuth angle of the UAV at tine $t$ with $0\leq\phi(t)\leq 2\pi$.

We consider sub-6 GHz band for interference limited downlink cellular network i.e., the presence of thermal noise power at a receiver is negligible compared to the interference power. We also assume that the MBSs and the UAV share common transmission bandwidth and full buffer traffic is used in every cell. Round robin scheduling algorithm is considered in all downlink transmissions. It is also assumed that the receivers can mitigate the Doppler spread stemming from the mobility of the UAV. 


\subsection{Path Loss Model}
\label{PL}
In this paper, we consider three different path loss models for the MBS-UAV, the MBS-UE, and the UAV-UE links as depicted in Fig.~\ref{back}. We choose them in a way to mimic the realistic cellular networks. Performance comparison of different path loss models for different links is provided in our previous work \cite{moin2}. 
\subsubsection{MBS-to-UE Link}
We consider Okumura-Hata path loss model (OHPLM) for modeling the path loss in MBS-to-UE links~\cite{hata}. This model is more relevant for rural and sub-urban environments where base-station height does not vary significantly. According to OHPLM, 
the path loss (in dB) observed at UE $k \in \mathcal{K}$ from MBS $m$ at time $t$ is given by:
\begin{equation}
 \xi_{{k,m}}(t)=A+B\log_{10}({d_{{k,m,t}}})+C.  
\end{equation}
 Here, ${d_{{k,m,t}}}$ is the Euclidean distance from MBS $m$ to user $k$ at time $t$. $A$, $B$, and $C$ are the factors dependent of the carrier frequency ${\text{f}_{\text{c}}}$ and antenna heights \cite{hata}. The parameters $A$, $B$, and $C$, in a suburban environment are given by~\cite{hata}:
\begin{align}
A &= 69.55+26.16\log_{10}({\text{f}_{\text{c}}})-13.82\log_{10}(\mathrm{h}_{\mathrm{bs}})-a(\mathrm{h}_{\mathrm{ue}}), \\ 
B &= 44.9-6.55\log_{10}(\mathrm{h}_{\mathrm{bs}}),\\
C &=-2\log_{10}({\text{f}_{\text{c}}}/28)^2-5.4,
\end{align}
 where ${\text{f}_{\text{c}}}$ is carrier frequency in MHz, $\mathrm{h}_{\mathrm{bs}}$ and $\mathrm{h}_{\mathrm{ue}}$ are the height of the MBS and the height of UE in meter unit, respectively. Since $\mathrm{h}_{\mathrm{bs}}$ and $\mathrm{h}_{\mathrm{ue}}$ are kept fixed, according to ~\cite{hata}, the parameters $A$, $B$, and $C$ will also be constant. The correction factor $a(\mathrm{h}_{\mathrm{ue}})$ due to UE antenna height can be defined as,
 \begin{equation}
     a(\mathrm{h}_{\mathrm{ue}})=[1.1\log_{10}({\text{f}_{\text{c}}})-0.7]\mathrm{h}_{\mathrm{ue}}-1.56\log_{10}({\text{f}_{\text{c}}})-0.8.
 \end{equation}

 For considering OHPLM, the carrier frequency (${\text{f}_{\text{c}}}$) should be in the range between 150 MHz and 1500 MHz, $\mathrm{h}_{\mathrm{bs}}$  between 30 m to 200 m, and $\mathrm{h}_{\mathrm{ue}}$ between 1 m to 10 m.
 \subsubsection{UAV-to-UE Relay Link}
 We deploy the mixture line-of-sight (LoS) and non-line-of-sight (NLoS) propagation model (MPLM) for modeling the path loss between UAV and the UEs\cite{mixturepl}. Since UAVs fly above the ground, they can get LoS channel with high probability. In fact, the higher a UAV goes, its LoS probability gets closer to one\cite{3gpp}. On the other hand, due to the presence of man-made structures on the ground, the link between UAV and UE can be in NLoS scenario, while the UAV is completing its mission. 
 
 According to MPLM, with a given horizontal distance, $z_{{k,u,t}}$,  between a UE $k$ and the UAV at time $t$, the LoS probability $\tau_L(z_{{k,u,t}})$ can be defined as\cite{mixturepl}:
 
 \begin{figure}[t]
 \label{backhaul}
\centering
\includegraphics[width=.85\linewidth]{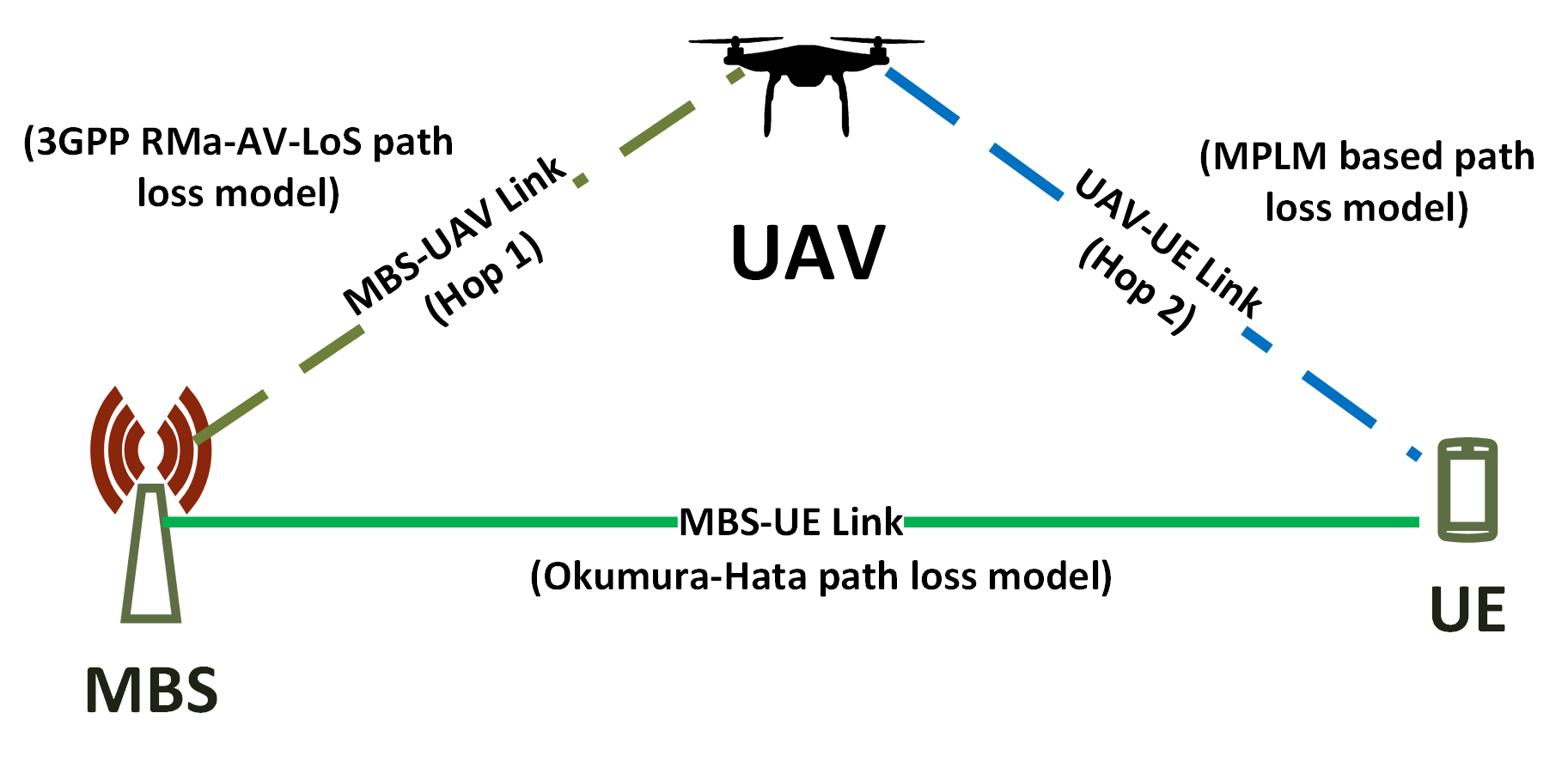}\\
\caption{{An illustration of using UAV as cellular network component with backhaul constraint. }}
\label{back}\vspace{-3mm}
\end{figure}
\begin{equation}
 \tau_L(z_{{k,u,t}})=\prod_{n=0}^{m} \bigg(1-\exp {\bigg\{-\frac {\mathrm{h}_{\mathrm{uav}}-(n+0.5)(\delta_{\mathrm{h}})}{2\hat{c}^2} \bigg\}}\bigg),
\end{equation}
where $m=\floor*{ \frac{z_{{k,u,t}}\sqrt{\hat{a}\hat{b}}}{1000}-1}$ and $\delta_{\mathrm{h}}=\mathrm{h}_{\mathrm{uav}}-\mathrm{h}_{\mathrm{ue}}$.
Here, a suburban area is defined as a set of buildings placed in a square grid in which $\hat{a}$ stands for fraction of the total land area occupied by the buildings, $\hat{b}$ is the mean number of buildings per sq. km, and the buildings height is defined by a Rayleigh PDF with parameter $\hat{c}$\cite{mixturepl}. Increasing these parameters will decrease the LoS probability with increasing elevation angle and thus, higher values of these parameters are used to characterize the LoS probabilities in urban, high-rise urban scenarios~\cite{mixturepl,mplm_param}. Consequently, the NLoS probability can be expressed as $\tau_N(z_{{k,u,t}})=1-\tau_L(z_{{k,u,t}}).$ 
After calculating the LOS and NLOS probabilities, we can get the path loss (in dB) at UE $k \in \mathcal{K}$ from the UAV at time $t$ as:
\begin{equation}
\begin{split}
 \xi_{{k,u}}(t)=& 10\log_{10}(\mathrm{P}_\textsubscript{uav} [(d_{k,u,t})^{-\alpha_{L}}\tau_L(z_{k,u,t})\\&+(d_{{k,u,t}})^{-\alpha_{N}}\tau_N(z_{{k,u,t}})]),   
\end{split}
\end{equation}
where $\mathrm{\alpha_{L}}$ and $\mathrm{\alpha_{N}}$ are the path loss exponents associated with the LoS path and the NLoS path, respectively. The values of $\hat{a}$, $\hat{b}$, and $\hat{c}$ along with path loss exponents  $\mathrm{\alpha_{L}}$ and $\mathrm{\alpha_{N}}$ are provided in Table \ref{mplm_param}.
\begin{table}[t]
\centering
\caption{MPLM parameters.} 
\scalebox{1.2}{
\begin{tabular}{cc} \hline
{\textbf{Parameter}} & {\textbf{Value}} \\ \hline

$\mathrm{\alpha_L}$ & 2.09\\ 
$ \mathrm{\alpha_N}$ &  3.75\\ 
$\hat{a}$ & 0.1\\ 
$\hat{b}$  & 100 \\ 
$\hat{c}$ & 10\\ \hline
\end{tabular}}
\label{mplm_param}
\end{table}
\subsubsection{MBS-to-UAV Link}
To model the path loss between an MBS and a UAV, we consider RMa-AV-LoS channel model specified in 3GPP\cite{3gpp}. According to \cite{3gpp}, the probability of LoS is equal to one if the UAV height falls between 40 m and 300 m. The instantaneous path loss (in dB) between an MBS, $m$ and the UAV can be expressed as:

\begin{equation}
\begin{split}
\xi_{{m,u}}(t)&=\text{max}\big(23.9-1.8\log_{10}(\mathrm{h}_{\mathrm{uav}}),20 \big)\log_{10}(d_{{m,u,t}})\\&+20\log_{10}\bigg(\frac{40\pi {\text{f}_{\text{c}}}}{3}\bigg),
\end{split}
\end{equation}
where  $\mathrm{h}_{\mathrm{uav}}$ is between 10 m to 300 m, while $d_{m,u,t}$ represents the 3D distance between the UAV and an MBS $m$ at time $t$. Note that even though fading will impact that the received signal quality at the receiver's end with increased height, the LoS probabilities associated the MBS-to-UAV links and the UAV-to-UE links will not be impacted due to presence of fading.

\section{Factors effecting the optimal UAV trajectory}
\label{sec:factors}
In this subsection, we will study the two important factors that can create significant impact on the optimal trajectories of a UAV acting as a relay. When a UAV starts flying in the sky, it needs to keep continuous communication with the cellular network via the backhaul links which can be provided by the MBSs nearby. While these MBS-to-UAV links are highly dependent on the antenna gain of the backhaul link, the end-to-end capacity of a relay depends on the quality of both MBS-to-UAV and  UAV-to-UAV links \cite{moin2}.
\subsection{Antenna Radiation Pattern}

We consider the 3GPP antenna radiation model to characterize the antenna radiation at the MBS. According to this model, each MBS is divided into three sectors as we have mentioned earlier, and each sector is equipped with eight cross polarized antennas $(\pm 45\degree)$, placed on a uniform linear array (ULA). Each of these antenna element pattern provide high directivity with a maximum gain in the main-lobe direction of about 8 dBi\cite{ant3gpp} and together they form antenna array that provides high gain towards the steering direction.
\subsubsection{Element Radiation Pattern}\label{radiation}
The 3GPP element pattern is realized according to the specifications in \cite{3gpp}, where the radiation pattern of each single cross polarized antenna element consists of both horizontal and vertical radiation patterns. These two radiation patterns $A_\text{{E,H}}(\phi^\prime)$ and $A_\text{{E,V}}(\theta^\prime)$ are obtained as \cite{1},\cite{2}:
\begin{figure}
		\centering
		\subfloat[]{
			\includegraphics[width=.75\linewidth]{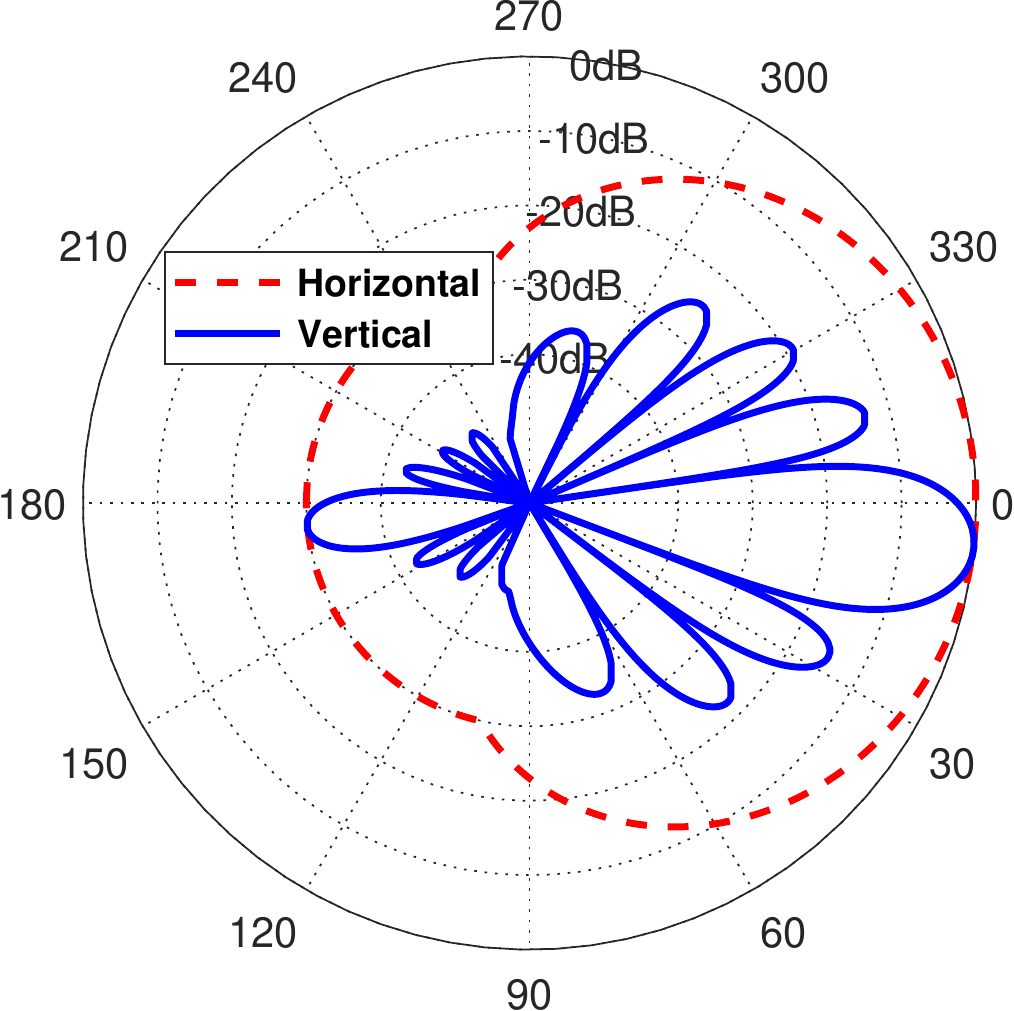}}
           		
		\subfloat[]{
			\includegraphics[width=.8\linewidth]{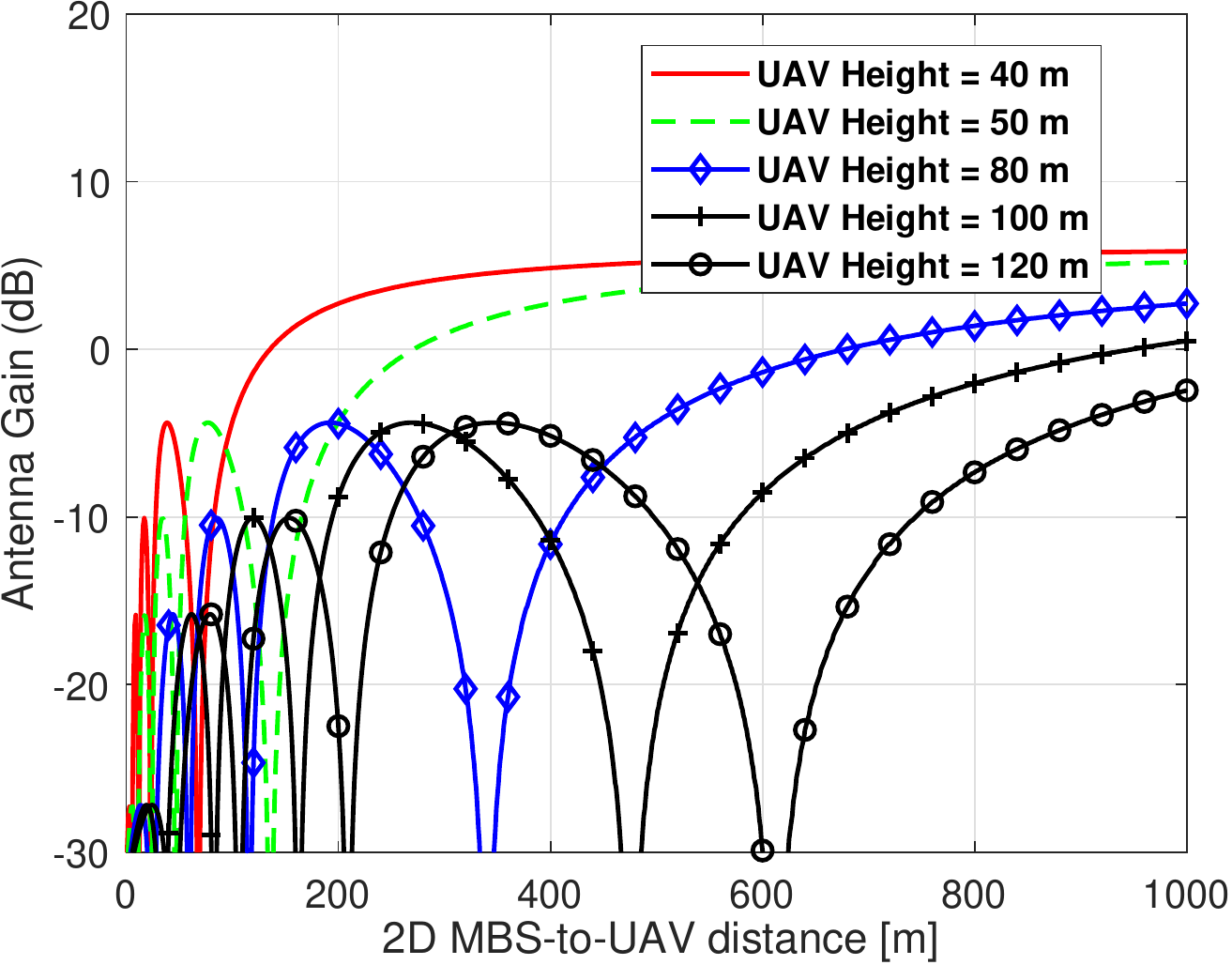}}
			
		\caption {
		{(a) Horizontal and vertical antenna pattern of an MBS after normalization consisting of a vertical array of 8 X-POL elements, each with 65\degree~half power beamwidth, downtilted by 6\degree.
		(b) Antenna gain between an MBS and a UAV aligned to the MBS’s horizontal bearing as a function of their 2D distance. Various UAV heights are considered.
        }}
		\label{antenaa_radiation}
\end{figure}

\begin{align}
   A_{\mathrm{E,H}}(\phi^\prime) &= - \text{min}\left\{ 12\left(\frac{\phi^\prime}{\phi^\prime_{3\mathrm{dB}}} \right)^2, \mathrm{A_m}\right\}, \\
   A_{\mathrm{E,V}}(\theta^\prime) &= - \text{min}\left\{ 12\left(\frac{\theta^\prime-90}{\theta^\prime_{3\mathrm{dB}}} \right)^2, \mathrm{SLA_V}\right\},
\end{align}

where $\phi^\prime_{3\mathrm{dB}}$ and $\theta^\prime_{3\mathrm{dB}}$ both refer to 3 dB beamwidth with same value 65\degree, $\mathrm{A_m}$ and $\mathrm{SLA_V}$ are front-back ratio and side-lobe level limit, respectively, with identical value 30 dB. Then, by combining together the vertical and horizontal radiation patterns of each element, we can compute the 3D antenna element gain for each pair of angles as:
\begin{equation}
A_{\mathrm{E}}(\theta^\prime,\phi^\prime)= G_{\mathrm{max}} - \text{min}\left\{-[A_{\mathrm{E,H}}(\phi^\prime)+A_{\mathrm{E,V}}(\theta^\prime)],\mathrm{A_m}\right\}.
\label{eq_element}
\end{equation}\par

The expression in \eqref{eq_element} provides the dB gain experienced by the UAV and the UEs with angle pair $(\theta^\prime, \phi^\prime)$ due to the effect of the 3D element radiation pattern.
\subsubsection{Array Radiation Pattern}
The antenna array radiation
pattern $A_{\mathrm{A}}(\theta^\prime,\phi^\prime)$ tells us how much power is radiated from an antenna array towards the steering direction $(\theta^\prime,\phi^\prime)$.
Following \cite{1},\cite{2}, the array radiation pattern with a given element radiation pattern $A_{\mathrm{E}}(\theta^\prime,\phi^\prime)$ from \eqref{eq_element} can be calculated as: 
\begin{equation}
  A_{\mathrm{A}}(\theta^\prime,\phi^\prime)=  A_{\mathrm{E}}(\theta^\prime,\phi^\prime)+\text{AF}(\theta^\prime,\phi^\prime,n).
\end{equation}

The term $\text{AF}(\theta^\prime,\phi^\prime,n)$ is the array factor with the number $n$ of antenna elements, given as:
\begin{equation}
\text{AF}(\theta^\prime,\phi^\prime,n)=10\log_{10}\big [1+\rho \big(|\mathbf{a}~.~\mathbf{w}^T|^2-1 \big)\big],
\end{equation}
where $\rho$ is the correlation coefficient, set to unity. The term $\mathbf{a}$ $\in \mathbb{C}^{n}$ is the amplitude vector, set as a constant
$1/\sqrt{n}$ while assuming that each antenna element has equal amplitude. The term $\mathbf{w}$ $\in \mathbb{C}^{n}$ is the beamforming vector, which includes the mainlobe steering direction, can be expressed as:

\begin{equation}
  \mathbf{w} = [w_{1,1}, w_{1,2}, . . . , w _{m_V,m_H}],  
\end{equation}
where~ $m_Vm_H=n$, $w_{p,r}=e^{j2\pi \big((p-1)\frac{\Delta V}{\lambda} \Psi_p +(r-1)\frac{\Delta H}{\lambda} \Psi_r \big)}$, \\$\Psi_p = \cos({\theta})-\cos({\theta_s})$, and $\Psi_r = \sin({\theta})\sin({\phi})-\sin({\theta_s})\sin({\phi_s})$. $\Delta V$ and $\Delta H$ stand for the spacing distances between the vertical and horizontal elements of the antenna array, respectively. We consider $\Delta V=\Delta H=\frac{\lambda}{2}$, where $\lambda$ represents the wavelength of carrier frequency. It is worth noting that the pair of angles $(\theta,\phi)$ is different from the steering pair $(\theta_s,\phi_s)$ where the main beam is steered due to beamforming. The mutual coupling effects is also omitted in our study.

Also note that, since we are considering a ULA along the $z$- axis, the array factor $\text{AF}(\theta^\prime,\phi^\prime,n)$ is only dependent on the vertical angle $\theta^\prime$. Similarly, $m_V$ will be equal to number of elements $n$ in equation (18) and $,m_H$ will be equal to one. Moreover, we consider analog beamforming technique by steering the main beam downtilted by $6^\circ$, and calculate beamforming vector $\mathbf{w}$ based on the locations of the static UEs as well the UAV. We also assume that the exact locations of the static UEs are known in prior. Beamforming vectors for different locations along the UAV trajectory were also calculated offline depending on the elevation and azimuth angle between an MBS and the UAV.

In Fig.~3(a), we show the normalized horizontal and vertical radiation pattern and in Fig.~3(b) we demonstrate how antenna gain between an MBS and a UAV
changes as a function of the 2D distance between them. As the horizontal distance between the BS and the UAV changes, the elevation angle (the angle between the $z$-axis and the line joining the UAV and the MBS) also changes which in turn changes the beamforming weight vector $\mathbf{w}$ in equation (18) and the single element vertical radiation pattern equation (14). These two will impact the total antenna array gain as specified in equation (16).

\subsection{Backhaul Constraint}
As mentioned earlier, it is required for the UAV to maintain command and control link with the core network for reliable and safe mission-critical UAV operation in BVLoS scenario\cite{backhaul}. On the other hand, while acting as BS in downlink scenario, UAV has to gather data from the core network through the backhaul link. Hence, we consider the UAV acting as a relay between the MBSs and the UEs in downlink and study the network performance. We assume amplify and forward (AF) technique, where the end-to-end signal-to-interference ratio (SIR) is calculated as the harmonic mean of the SIRs related to UAV-UE link and MBS-UAV link\cite{chetan},\cite{sir}. An example of using UAV as a relay in downlink scenario is depicted in Fig.~\ref{back}. Let the SIRs associated with the MBS-UAV link be denoted as $\gamma_\textsuperscript{mbs-uav}$ and $\gamma_\textsuperscript{uav-ue}$ for a UE $k$. According to \cite{sir}, the end-to-end SIR of UE $k$ can be calculated as:

\begin{equation}
    \gamma_k=\frac{2\gamma_\textsuperscript{mbs-uav} \times \gamma_\textsuperscript{uav-ue}}{\gamma_\textsuperscript{mbs-uav}+\gamma_\textsuperscript{uav-ue}}.
\end{equation}
Note that $\gamma_\textsuperscript{mbs-uav}$ is dependent on both (1) antenna radiation pattern of the MBS antennas and (2) 3GPP path loss model for MBS-to-UAV link discussed in the previous subsections. The equations for calculating SIRs are specified in Section V.
\section{UAV trajectory Optimization Problem}
\label{sec:prob}
Considering the system model described in Section II, the received power at user $k$ from  MBS $m$  at time $t$, can be calculated as ${S_{{m},t}}=\frac{{\mathrm{P}_{\mathrm{mbs}}}}{10^{\xi_{{k,m}}(t)}/10}$. Similarly, the received power at user $k$ from  the UAV at time $t$, can be calculated as ${S_{{u,t}}}=\frac{{\mathrm{P}_{\mathrm{uav}}}}{10^{\xi_{{k,u}}(t)}/10}$. Here, $\xi_{{k,m}}(t)$ and $\xi_{{k,u}}(t)$ are path losses in dB which are calculated according to path loss models discussed in Section \ref{PL}. During each $t$, a UE connects to either its nearest MBS or the UAV, whichever provides the best signal-to-interference ratio (SIR)\cite{merwaday}. Assuming round-robin scheduling, we can express the spectral efficiency (SE) in bps/Hz unit of user $k$ at time $t$ using Shannon's capacity as follows: 
\begin{align}
{C_{k}(t)}=\frac{\log_{2}(1+{\gamma_{k}(t))}}{{N_{\mathrm{ue}}}},
\label{eqn:rate}
\end{align}
where ${\gamma_{k}(t)}$  is the instantaneous SIR of $k$-th user at time $t$ and ${N_{\mathrm{ue}}}$ is the number of users in a cell. Then ${\gamma_{k}(t)}$ can be expressed as:
\begin{align}
\gamma_{k}(t)=\frac{S_{{i},t}}{\sum_{j \neq i} S_{{j},t}},
\end{align}
where $S_{{i},t}$ is the received power at user $k$ from  transmitter (MBS/ UAV) $i$,  with which the user $k$ is associated at time $t$. The instantaneous sum-rate of the network at time $t$ can be expressed as follows:
\begin{equation}
    C(t)= \sum_{k=1}^K {C_{k}(t)},
\label{eq}
\end{equation}where $C_{k}(t)$ is the SE of user $k$ at time $t$ as presented in~(\ref{eqn:rate}). It is worth noting that in our work, the UAV and the UEs are equipped with omnidirectional antennas. According to~\cite{Balanis}, the polarization direction of an omnidirectional antenna can be vertical, horizontal, circular, or dual, and the SIR values will be degraded due to the polarization mismatch in the MBS-to-UE link or in the UAV-to-UE link~\cite{mmimo}. However, we did not consider the impact of polarization on the SIR values in this paper.

Now, using \eqref{eq} and the UAV mobility in Section III A, we can formulate our 3D trajectory optimization problem over the total mission duration of the UAV as follows:  

\begin{maxi!}
	  {\mathbf{r(t)}}{\frac{1}{T}\int_{t=0}^{T}C(t)\;dt}{}{}\label{prob1}
	  \addConstraint{\sqrt{\dot{x}(t)^2+\dot{y}(t)^2+\dot{z}(t)^2}}{\leq {V_{\mathrm{max}}}, } \; { t \in [0,T]}\label{prob2}
	  \addConstraint{[x(0),y(0),z(0)] =}{[x_{\rm s},y_{\rm s},z_{\rm s}]}{}\label{prob3}
	  \addConstraint{[x(T),y(T),z(T)] =}{[x_{\rm f},y_{\rm f},z_{\rm f}]}{}\label{prob4}
	  \addConstraint{\hspace{-2cm} h_{\mathrm{min}} \leq}{z(t)}{\leq h_{\mathrm{max}}}.\label{prob5}
\end{maxi!}

Here, (\ref{prob2}) ensures that the velocity of the UAV does not exceed the maximum limit V\textsubscript{max}, while (\ref{prob3}) and (\ref{prob4}) fix the initial and final location of the mission, respectively. Finally, (\ref{prob5}) represents the UAV height constraint.
The trajectory generated from the above optimization problem aims at maximizing the time-averaged sum-rate of the users over the whole mission duration. Here, we also consider $T \geq T_\textsubscript{min}$, where $T_\textsubscript{min}$ is as in (\ref{eqn:tmin}), so that there exists at least one feasible solution to the above optimization problem. Note that in this paper, we did not consider any boundary conditions or path constraints which might affect the feasible set of the problem.

The optimization problem in (23) is very difficult to solve efficiently since the objective function is non-convex with respect to $\mathbf{r(t)}$. The search space of the problem can also be large and continuous. Hence, we use approximate DP based technique for obtaining high-quality close to optimal solutions while being computationally efficient.
\section{Dynamic Programming for 3D Trajectory Optimization}
\label{sec:dp}
In this section, using discrete-time approximation of (23), we formulate it as finding an optimal control of a discrete dynamical system. Then we will use DP technique to find the approximate optimal policies/trajectories of this discrete optimal control problem. We choose DP due to its ease of implementation and ability to deliver low-complexity sub-optimal solutions that are close to the optimum solution. To tackle the curse of dimensionality inherent in discrete optimal control problem~\cite{Bellman:1957}, approximate dynamic programming~\cite{adp} can be used. We plan to consider this in our future work.\footnote{It is worth noting that our problem can
also be solved by using general nonlinear programming algorithms and other off-the-shelf solvers
such as CPLEX, BARON, etc. In our future works, we will consider comparing our DP-based
solution with other commercial or open-source solvers.}
Note that, in our case, mission duration $T$ is finite and hence, this is a Finite horizon DP problem \cite{dp}.

At first, the optimization problem in (23) is discretized to obtain approximation of the optimal trajectories. The time period $[0,T]$ is divided into $N$ equal intervals of duration $\delta=T/N$ and is indexed by $i=0,....,N-1$. The value of $N$ is chosen so that UAV's position, velocity, and heading angle can be considered constant in an interval. The rate of UE $k$, ${R_{k}(i)}$, at time interval $i$, will be dependent on the position of the UAV along the 3D Cartesian coordinates at that particular time interval. After the discretization, we can write the discrete-time dynamic system as follows:
\begin{equation}
\label{eqn:dp}
    \mathbf{r}_{i+1}=\boldsymbol{r}_{i}+ f(i,\boldsymbol{r}_{i},\boldsymbol{u}_{i}),\quad  i=0,1,...,N-1,
\end{equation}
where $\boldsymbol{r}_{i}=[x_i \; y_i \; z_i]^T$ is the state or the position of the UAV at time $i$ and $\boldsymbol{u}_{i}=[v_i \;\underline{\phi_i} \; \underline{\theta_i}]^T$ stands for the control action i.e., velocity $v_i$, azimuth angle $\underline{\phi_i}$, and elevation angle $\underline{\theta_i}$ respectively, in the $i$-th time interval. By taking control action at each interval $i$, the UAV will move to next state for taking that specific control action. Starting with initial state $\boldsymbol{r}_{0}=[x_{\rm s}, y_{\rm s}, z_{\rm s}]$, the subsequent states can be computed by adding $f(i,\boldsymbol{r}_{i},\boldsymbol{u}_{i})$ with the current state. We can compute the state transition vector $f(i,\boldsymbol{r}_{i},\boldsymbol{u}_{i})$ as:
\begin{equation}
    f(i,\boldsymbol{r}_{i},\boldsymbol{u}_{i})= \begin{pmatrix}
 v_i\sin\underline{\theta_i}\cos\underline{\phi_i}  \\
 v_i\sin\underline{\theta_i}\sin\underline{\phi_i}\\
  v_i\cos{\underline{\theta_i}} \\
\end{pmatrix}.
\end{equation}

Let $\boldsymbol{\pi} = \{\boldsymbol{u}_{0},\boldsymbol{u}_{1}, . . . ., \boldsymbol{u}_{N-1}\}$ be a set of sequential decisions for reaching the final state, $\boldsymbol{r}_N=[x_{\rm f}, y_{\rm f}, z_{\rm f}]^T$ by starting from the initial state, $\boldsymbol{r}_{0}$. Further let the total cost function of using $\boldsymbol{\pi}$ with the initial state be as:
\begin{equation}
    J_{\pi}(\boldsymbol{r}_0)=J_{\boldsymbol{r}_N}+  \sum_{i=0}^{N-1}\sum_{k=1}^K {C_{k}(i)},
\end{equation}where the terminal cost $J_{\boldsymbol{r}_N}$ is the cost when the UAV reaches the position $[x_{\rm f}, y_{\rm f}, z_{\rm f}]$, which can be expressed as follows:
\begin{equation}
\label{dp_goal}
   J_{\boldsymbol{r}_N}= \sum_{k=1}^K {C_{k}(N)},
\end{equation}
\begin{algorithm}[!t]\small 
	\caption{3D Trajectory Optimization using DP}
    \label{alg:Alg2}
	\begin{algorithmic}[1]
		\STATE \textbf{Input:} $L_{\rm s}$, $L_{\rm f}$, $N$
		\STATE  Divide the network into $m$ number of discrete 3D grid points (states) and start from the destination point $L_{\rm f}$
		\STATE Initialize: matrix $\mathbf{A}$ of size ($N-2 \times m$) with zeros
		\STATE Calculate $J_{\boldsymbol{r}_N}$ by using \eqref{dp_goal} and save it for recursion
		\STATE  \textbf{for}  $t=N-1 \;to\;  1$ \textbf{do}
		\STATE  \hspace{0.3cm} \textbf{for}  $ii=1 \;to\;  m$ \textbf{do}
		\STATE  \hspace{0.6cm} Calculate the sum-rates of all allowed neighbor states \\ \hspace{0.6cm} which can be reached from current state $ii$
		\STATE  \hspace{0.6cm} \textbf{if} sum-rate of a neighbor state was not saved previously
		\STATE  \hspace{1cm} set sum-rate of that neighbor state equal to $-\infty$
		\STATE  \hspace{0.6cm} \textbf{end if}
		\STATE \hspace{0.6cm} Add the sum-rate of each neighbor state with that of \\ \hspace{0.6cm} current state $ii$ and save for recursion
		\STATE  \hspace{0.6cm} By using \eqref{dp_iter}, find the control action which maximizes \\ \hspace{0.6cm} the current total sum-rate and save it at $\mathbf{A}(t,ii)$
		\STATE \hspace{0.3cm} \textbf{end for}
		\STATE  \textbf{end for}
		
		\STATE Set \textit{current point} = $L_{\rm s}$
		\STATE \textbf{for}  $jj=1 \; to \; N-2$ \textbf{do} 
		\STATE \hspace{0.3cm} $traject(jj) = \mathbf{A}(jj,\textit{current point})$
		\STATE \hspace{0.3cm} $\textit{current point}= traject(jj)$ \STATE \textbf{end for}
		\STATE ${\boldsymbol{\pi}}^*=(L_{\rm s}+traject+L_{\rm f})$
		\STATE \textbf{Output:} ${\boldsymbol{\pi}}^*$
	\end{algorithmic}
\end{algorithm}
Note that, the mission ends when the UAV reaches the final state, $\boldsymbol{r}_N$ and hence, we compute it separately by setting the UAV location at $[x_{\rm f}, y_{\rm f}, z_{\rm f}]$. By using DP, our overall goal is to find an optimal policy or sequence of decisions with an aim to optimize the time-averaged total sum-rate. An optimal policy vector $\boldsymbol{\pi}^*$, starting from $\boldsymbol{r}_0$ is given by: 
\begin{equation}
\label{dp_policy}
 {\boldsymbol{\pi}}^*= \max_{\boldsymbol{\pi} \in \Pi} J_{\pi}(\boldsymbol{r}_0),
\end{equation}	
where $\Pi=\{\boldsymbol{u}_{i},\forall i=0,1,...,N-1 |v_i\leq V_\textsubscript{max},0\leq \underline{\phi_i} \leq 360^{\degree}, 0\leq \underline{\theta_i} \leq 180^{\degree}\}$. In other words, $\Pi$ is the set of all possible control actions which provide routes towards subsequent states as presented in (\ref{eqn:dp}).

In order to find the optimal policy ${\boldsymbol{\pi}}^*$ for \eqref{dp_policy}, we use  Bellman's theorem. According to Bellman's principle of optimality (PO) \cite{dp}, an optimal trajectory for a discrete decision problem going from $i=0$ to $i=N-1$ is also optimal for the subproblem going from $i=n$ to $i=N-1$, where $0 \leq n \leq N-1$. Let $\boldsymbol{\pi}^*= \{\boldsymbol{u}_{0}^*,\boldsymbol{u}_{1}^*, . . . ., \boldsymbol{u}_{N-1}^*\}$ be an optimal policy of the optimization problem (\ref{dp_policy}). Now, consider the tail subproblem whereby the UAV starts from state $\boldsymbol{r}_{n}$ at time instance $n$ and we want to calculate the optimal trajectory from $\boldsymbol{r}_{n}$ to the goal state $\boldsymbol{r}_{N}$. According to PO, the tail segment of optimal policy $\boldsymbol{\pi}^*$, $\{\boldsymbol{u}_{n}^*,\boldsymbol{u}_{n+1}^*, . . . ., \boldsymbol{u}_{N-1}^*\}$ will maximize the total sum-rate for the path planning problem where the UAV starts from $\boldsymbol{r}_{n}$ and reaches $\boldsymbol{r}_{N}$ within time $N-n$.

Based on the PO, DP proceeds backward in time from $N-1$ to 0. The optimization problem presented in (\ref{dp_policy}), can be solved recursively using Bellman's equations by moving backwards in time as follows~\cite{dp}, \cite{rajeev}:
\begin{equation}
\label{dp_iter}
    J(\boldsymbol{r}_{i})=\max_{\boldsymbol{u}_{i}} \left\{\sum_{k=1}^K {R_{k}(i)}+ J(\boldsymbol{r}_{i+1})\right\},\quad  i=N-1,..,0.
\end{equation}

The solution of the optimization problem (\ref{dp_policy}), maximizes (\ref{dp_iter}). Note that the terminal cost is calculated using (\ref{dp_goal}). The subsequent steps from $i=N-2$ to $i=0$ can  be calculated using (\ref{dp_iter}). Since for each state, we have to calculate $v_i$, $\underline{\phi_i}$, and $\underline{\theta_i}$, this solution is still computationally expensive. We present a pseudocode of the proposed DP based 3D trajectory optimization in Algorithm \ref{alg:Alg2}. 

\textbf{Proposition 1:} \textit{Let us consider a 3D rectangular cuboid network of length $l$, width $w$, and height $h$. Let $l$ and $w$ lie on $x$ and $y$ axis, respectively and they meet on the origin $(0,0,0)$. If there exists a unique position of UAV ($x^*,y^*,z^*$) which maximizes the instantaneous sum-rate $C(t)$ and the UAV's start and destination points are $(0,0,h)$ and $(l,w,h)$, respectively, then the UAV will be able to visit ($x^*,y^*,z^*$) if $T\geq \frac{\sqrt{w^2+h^2}+\sqrt{l^2+h^2}}{V_{\text{max}}}$.}

\begin{proof}
Assume that, the optimal point lies on the line connecting the vertices $(0,0,h)$ and $(l,w,h)$. Since we consider $T\geq T_{\text{min}}$, according to (1), the optimal trajectory will be able to pass through the optimal point. Now consider another scenario where ($x^*,y^*,z^*$) coincides with the point $(0,w,0)$. Then, the UAV have to fly with maximum velocity to reach $(0,w,0)$ in order to maximize the sum-rate \cite{rajeev} and then fly again with $V_{\text{max}}$ to reach the destination $(l,w,h)$. With simple algebraic steps, we can  show that this can only be possible if $T\geq \frac{\sqrt{w^2+h^2}+\sqrt{l^2+h^2}}{V_{\text{max}}}$. Similar arguments can also be presented, if the ($x^*,y^*,z^*$) coincides with the point $(0,l,0)$. For any $T< \frac{\sqrt{w^2+h^2}+\sqrt{l^2+h^2}}{V_{\text{max}}}$, the UAV will not be able to visit the optimal point due to mission duration constraint.
\end{proof}

\textbf{Proposition 2:} \textit{If the mission duration, $T$ is greater than the quantity $\frac{\sqrt{w^2+h^2}+\sqrt{l^2+h^2}}{V_{\text{max}}}$, then for a 3D rectangular cuboid network like above, the UAV will fly with $V_{\text{max}}$ except at the point ($x^*,y^*,z^*$). }
\begin{proof}
This can be proved by following the proof of \cite[Prop. 2]{rajeev}. Since $T > \frac{\sqrt{w^2+h^2}+\sqrt{l^2+h^2}}{V_{\text{max}}}$, according to Remark 2, the optimal trajectory will pass through ($x^*,y^*,z^*$). Let us assume the segment of the optimal trajectory, which does not encompass the optimal point ($x^*,y^*,z^*$), to be $Z$. Let us also consider two maximum UAV velocities, $v_1=V_{\text{max}}$ and $v_2<V_{\text{max}}$. The the total sum-rates associated with these velocities can be calculated as:

\begin{align}
   C_1(t)=\int_{Z,v_1}C(t)\; dt,\\
   C_2(t)=\int_{Z,v_2}C(t)\; dt.
\end{align}

Since $v_2<v_1$, the UAV will spend $\delta_t>0$ time on the optimal point for maximum UAV velocity, $v_1$ and so, the total sum-rate of the network will be increased by $C^*\delta_t$, where, $C^*$ is the sum-rate achieved when the UAV visits ($x^*,y^*,z^*$). Therefore, the optimal trajectory will fly over $Z$ with maximum velocity, $V_{\text{max}}$ to save time which can be utilized to hover over the optimal point.
\end{proof}
    
\begin{figure}[t]
\centering
\includegraphics[width=0.85\linewidth]{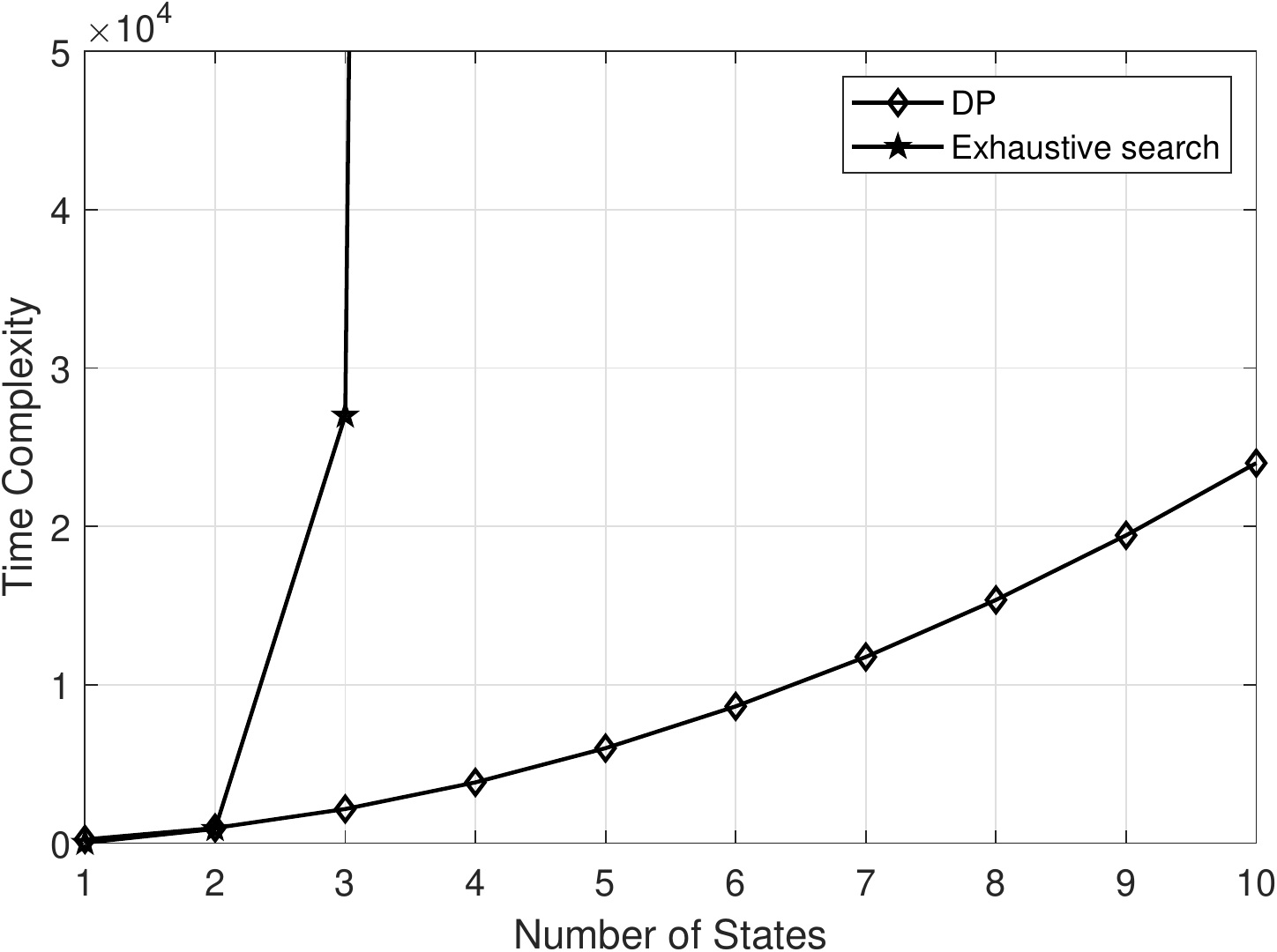}
\caption{Time versus complexity comparison between DP and ES algorithm for  $N = 30 $.}
\label{dp_exhaus}
\end{figure}

\subsection{Complexity Analysis}
Without stochastic influence,  discrete DP problems face the curse of dimensionality \cite{Bellman:1957}. The number of states grows with the dimensions and along with grid resolution, which makes the DP computation and memory requirement expensive. In the case of optimal path planning, an increase in the number of discrete time segments $N$ causes an increase in the number of combinations of discrete states to be examined at each time instance. The time complexity of the DP algorithm will be $\mathcal{O}(Nas^2)$, where $N$ stands for the number of discrete time slots, $a$ represents number of possible actions at a certain state, and $s$ stands for the number of unique states/grid points along the 3D Cartesian coordinate. Also, the number of grid points increases exponentially with the grid size resolution as mentioned above. On the other hand, for a similar setup, an exhaustive search (ES) algorithm will require to compute $N^s $ trajectories to find the optimal one, since at each time instance, the UAV could be at one of $s$ possible states. Hence, the time-complexity of the ES algorithm will be $\mathcal{O}(N^s)$. Fig.~\ref{dp_exhaus} shows the time complexities of DP and ES for $T= 240$ s which translates into $N=30$. It can be observed that the complexity of the DP increases quadratically with an increasing number of grid points whereas the ES complexity grows exponentially.  

\section{Numerical Results}
\label{sec:simulation}
\begin{table}[t]
\centering
\renewcommand{\arraystretch}{1.1}
\caption{Simulation parameters.} 
\scalebox{1}{
\begin{tabular}{cc} \hline
{\textbf{Parameter}} & {\textbf{Value}} \\ \hline
P\textsubscript{mbs} & 46 dBm  \\ 
P\textsubscript{uav} & 30 dBm \\ 
antenna downtilt angle & $6^\circ$ \\ 
V\textsubscript{max} & 18.75 m/s \\ 
$[x_{\rm s},y_{\rm s},z_{\rm s}]$ & [0, 0, 0.04] km \\ 
$[x_{\rm f},y_{\rm f},z_{\rm f}]$ & [1, 1, 0.04] km\\ 
h\textsubscript{min} & 40 m\\ 
h\textsubscript{max} & 120 m\\ 
h\textsubscript{bs} & 30 m\\ 
h\textsubscript{ue} & 2 m\\ 
${{\text{f}_{\text{c}}}}$ & 1.5 GHz\\ 
$\lambda$\textsubscript{mbs} & 2, 3,  4 per $\text{km}^2$\\ 
$\lambda$\textsubscript{ue} & 20, 50, 100 per $\text{km}^2$\\ 
Traffic Model & Full buffer\\ 
Scheduler & Round-robin\cite{roundrobin}\\ \hline
\end{tabular}}
\label{tab}
\end{table}
In this section, we provide numerical results to test the effectiveness of the proposed DP approach in Section VI. We numerically obtain the optimal trajectories of the UAV by applying DP and use Matlab simulations to analyze the effects of antenna radiation patterns on the optimal trajectories of the UAV acting as a relay. The UEs and the MBSs are distributed in an area of 1 $\times$ 1 $\text{km}^2$, whereas the UAV can fly over an area of 1.2  $\times$ 1.2 $\text{km}^2$. Such extension allows us to study the UAV behavior for antenna radiation and backhaul constraint more explicitly. The positions of the MBSs and the UEs are determined by uniform random distribution with density parameters $\lambda\textsubscript{mbs}$ and $\lambda\textsubscript{ue}$ per $\text{km}^2$, respectively. We run simulation for 1000 realizations and then compute the average to study the performance of the optimal 3D trajectories. The antenna downtilt angle is considered as 6\degree. Unless otherwise specified, the simulation parameters and their default values are listed in Table \ref{tab}.These values are chosen to reflect a realistic
UAV communication scenario, where the UAV traverses over an interference limited downlink cellular
network situated in a suburban area.
\begin{figure}[h!]
		\centering
		\subfloat[]{
			\includegraphics[width=.85\linewidth]{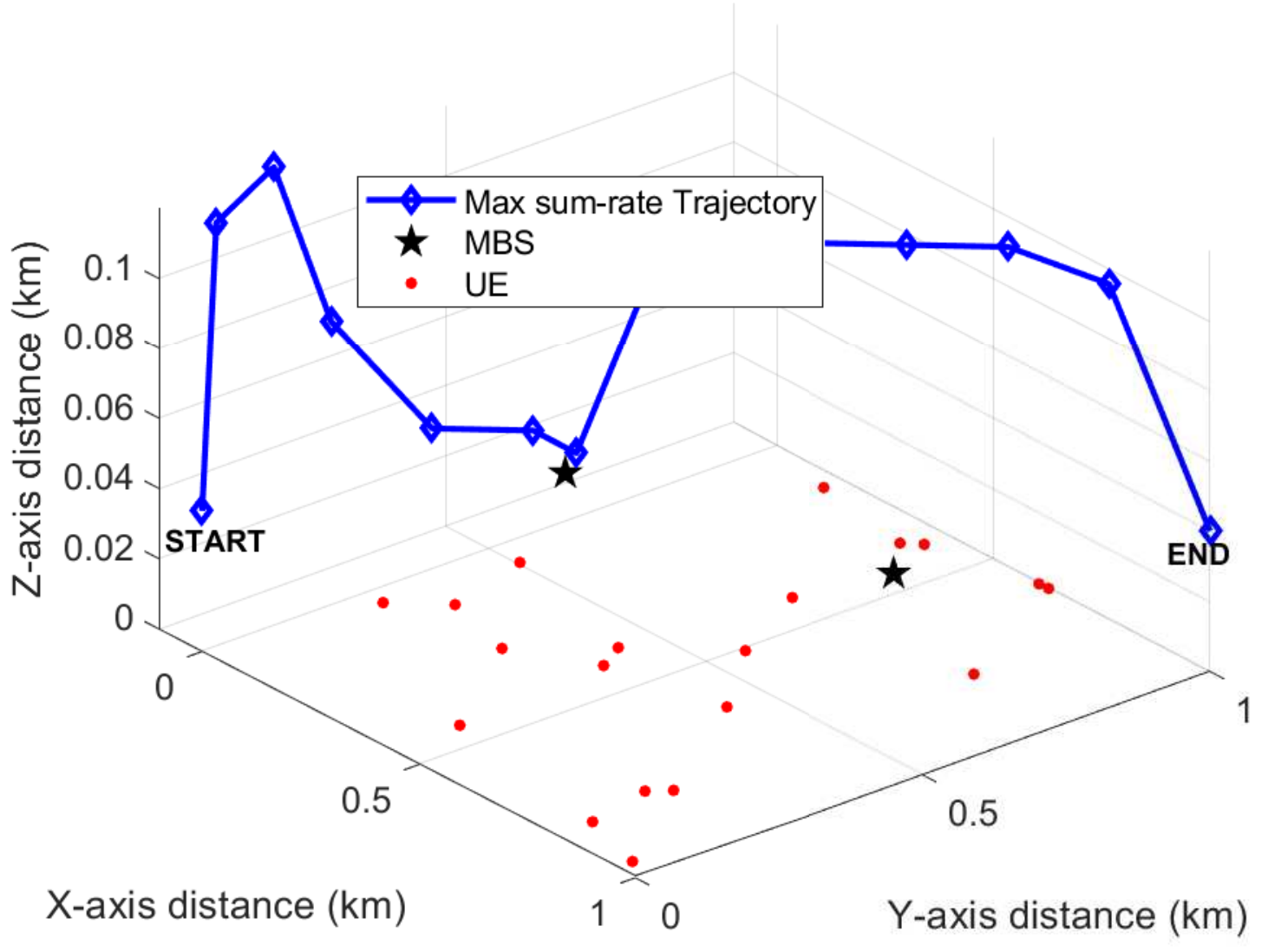}}
           		
		\subfloat[]{
			\includegraphics[width=.85\linewidth]{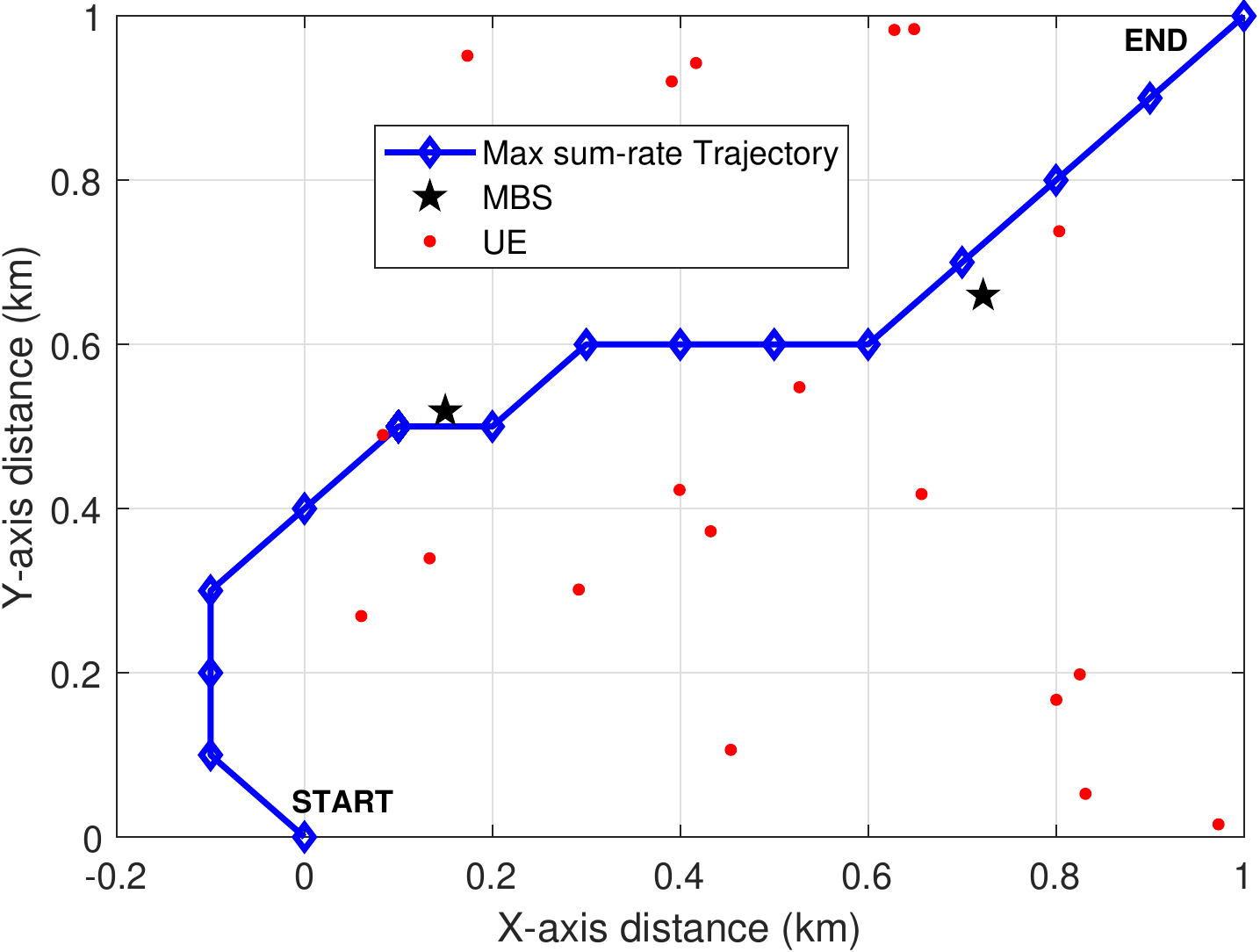}}
			
		\caption {
		{(a) Optimal trajectory with $N_{\text{MBS}}=2$ and 20 UEs. (b) Horizontal projection of the 3D trajectory on $xy$ plane.
        }}
		\label{traj_3D}
\end{figure}

For reducing computational complexity, the possible UAV 3D locations or states are divided into discrete segments. Due to the height restriction imposed by the Federal Aviation Administration (FAA) for small commercial UAVs\cite{faa}, we assume that the UAV can change its altitude from 40 m to 120 m with 10 m steps. We also discretize the square map $\mathcal{A}$ into steps of 100 m, resulting in total 1521 unique states or positions (169 unique grid positions in $xy$ plane and nine positions along the $z$ axis for each of the 169 grid positions). We also discretize the time into $\delta= 8$ seconds. Since we have segmented all possible states into finite discrete geometrical positions, we consider the following control actions on the map:
\begin{equation}
\label{act}
    \boldsymbol{u}_{i} \in \Bigg\{\begin{bmatrix} 0~\text{m/s}\\0\\ 0 \end{bmatrix},\begin{bmatrix} 13.98~\text{m/s} \\\mathbf{\phi}\\ \theta  \end{bmatrix},\begin{bmatrix} 18.75~\text{m/s} \\\mathbf{\phi} \\ \theta+\frac{\pi}{4} \end{bmatrix} \Bigg\},
\end{equation}
where,
\begin{equation}
\theta \in \left\{0,\frac{\pi}{2},\pi,\frac{3\pi}{2} \right\},
\end{equation}
\begin{equation}
\phi \in \text{tan}^{-1}\left\{ -0.5,-0.4,...,0.4,0.5\right\}.
\end{equation}
Actions in (\ref{act}) imply that the UAV either can stay at its current location or can change its height by keeping $x$ and $y$ coordinate points fixed. It can also move towards one of the 72 discrete neighbor points (eight discrete grid positions in $xy$ plane separated by $45^{\circ}$ and nine points along the $z$ axis), from a state. {Examples of such control specifications can also be found in~\cite{rajeev,gesbert}.}

\subsection{Optimal Trajectories}
In the following, we investigate the trajectories associated with the max sum-rate as in (23), in a network with 2 MBSs and 20 UEs for $T=240$~s, starting from start point $(0,0,0.04)$~km to destination $(1,1,0.04)$~km. In Fig.~\ref{traj_3D}(a), we plot optimal trajectory for a random realization. We also show the corresponding 2D plot on the $xy$ plane in Fig.~\ref{traj_3D}(b) for reader's convenience. It can be observed from both figures that the optimal trajectory tends to move towards grid points not too close to the MBSs in order to associate a few UEs and provide downlink coverage. Another interesting observation is that, while completing the mission, the UAV tends to reach the optimal point (highest value among the 1521 points) quickly and hover there for a while before it starts moving towards the final destination to meet the time constraint. This point is (0.1, 0.5, 0.04)~km for the optimal trajectory and this phenomenon is consistent with Proposition 2. At the optimal point, the UAV reaches the lowest allowed height (40 m) in order to decrease the path loss in the UAV-to-UE link. The UAV trajectory is also dependent on the mission duration $T$. For small $T$, the UAV tends to follow a straight line path since it will have less freedom to roam around the network, whereas large $T$ will allow the UAV to spend more time on the optimal point for increasing the sum-rate of the UEs. The UAV also has to maintain the backhaul link and hence, it does not go to the furthest grid points. In other words, the UAV has less freedom to hover over the area due to the backhaul constraint pertinent to the MBS-to-UAV link\footnote{UAV can not fly precisely over each point on the optimal discrete trajectory due to kinematic constraint. Bezier curves can be used to smooth the generated paths as illustrated in \cite{moin2,moin}}.  

\begin{figure}[t] 
	\centering
	\subfloat[]{
		\includegraphics[width=.85\linewidth] {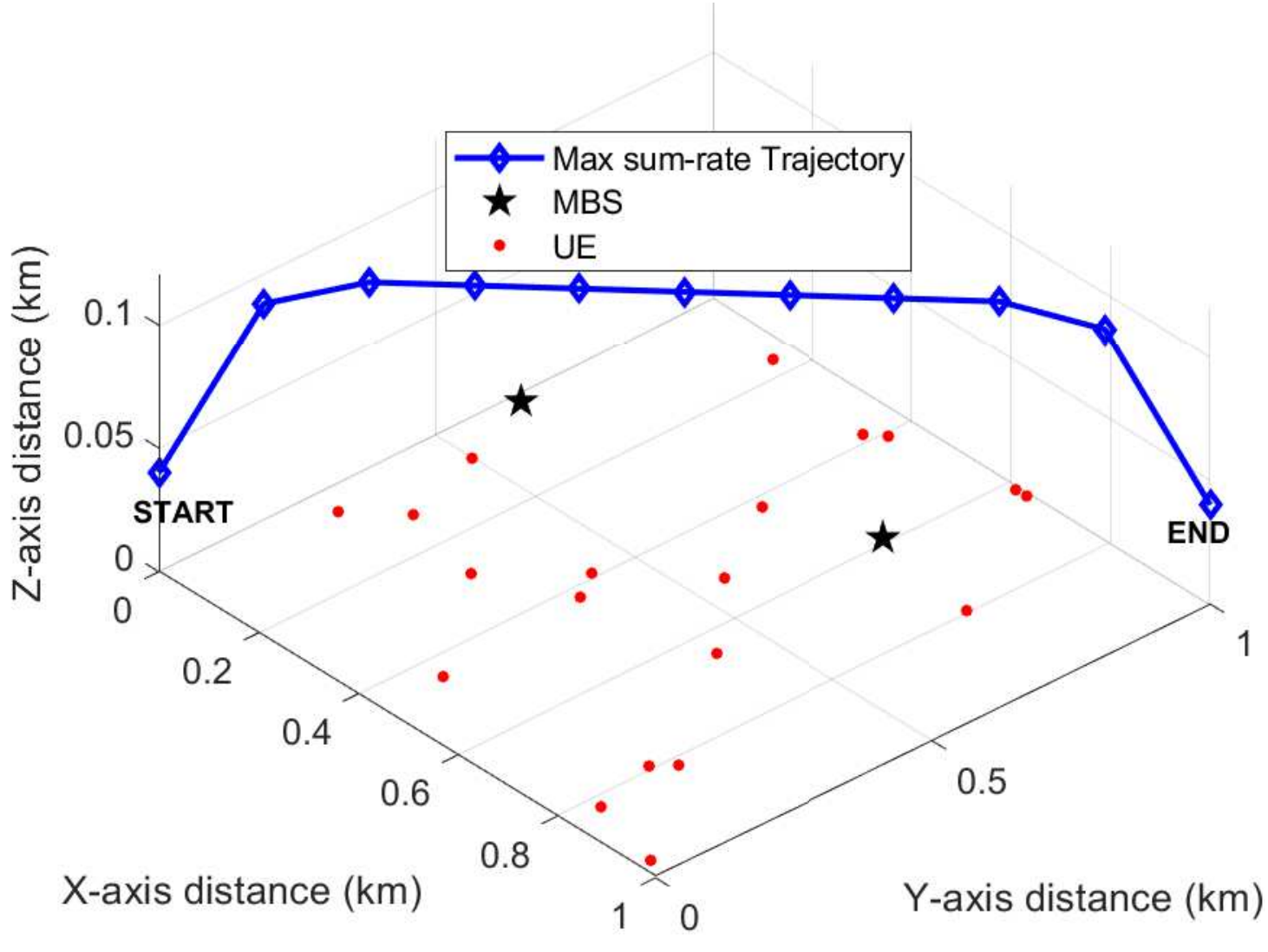}
		\label{small}}
		
	\subfloat[]{
		\includegraphics[width=.85\linewidth] {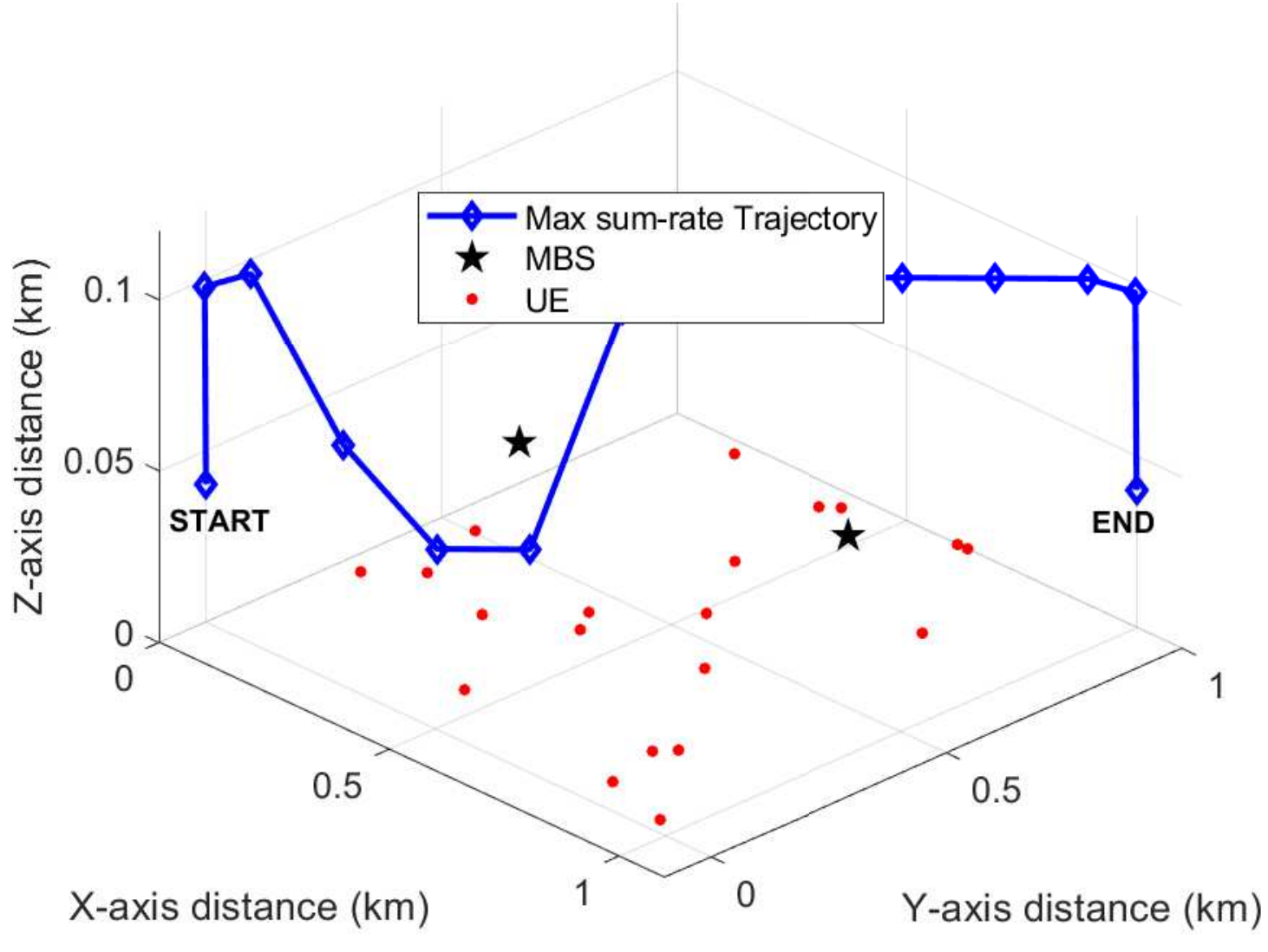}}
		
	\subfloat[]{
		\includegraphics[width=.85\linewidth] {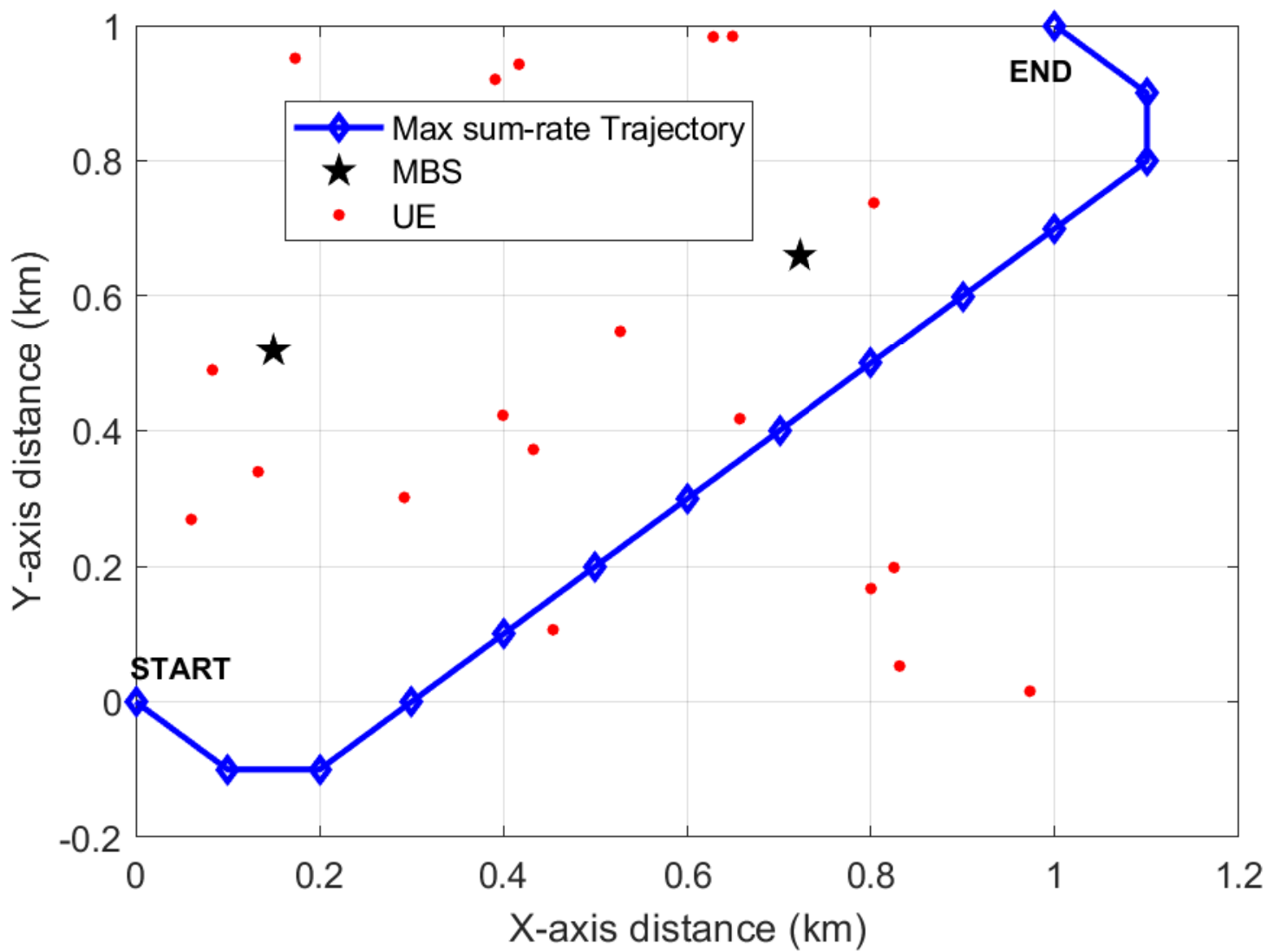}}
	\caption{ Impact of $T$ on the optimal UAV trajectory with $N_{\text{MBS}}=2$ and 20 UEs. (a) Optimal UAV trajectory with $T=80$~s. (b) Optimal UAV trajectory with $T=120$~s. (c) Horizontal projection of the 3D trajectory with of $T=120$~s on $xy$ plane.}
	\label{traj_3D_response_new}
\end{figure}

Effect of $T$ on the optimal trajectories is shown in Fig.~\ref{traj_3D_response_new} and after observing Fig.~\ref{traj_3D_response_new}(a), we can conclude that for $T=80$~s, the UAV will be forced to follow a straight line trajectory for completing its primary mission since it will have less freedom to roam around the network. If we increase the mission duration to $120$~s, the UAV will get more chances to explore the network for increasing the sum-rate. The 3D and 2D trajectories for $T=120$~s are depicted in Fig.~\ref{traj_3D_response_new}(b) and Fig.~\ref{traj_3D_response_new}(c), respectively. We also obtain the optimal trajectory associated with $T=400$~s which is the same path as shown in Fig. 5. The UAV will tend to reach the optimal point and hover there for a while before it starts moving towards the final destination. The only difference is that the UAV spends the extra time hovering over the optimal point for increasing the capacity. This observation is in line with our analysis presented in Proposition 2.

\subsection{Spectral Efficiency Comparison}
To study the network performance of 3D optimal trajectories, we first generate 1000 random networks for $\lambda_{\text{mbs}}=$ 2, 3, and 4 per $\text{km}^2$. Then we calculate the time-averaged total SE (bps/Hz) of the networks using \eqref{prob1} and divide it by the number of the total UEs of the network to get per UE SE (bps/Hz/UE). We also compute the time-averaged per UE SE for those networks without considering any UAV. 
\begin{figure}[t]
\centering
\includegraphics[width=0.85\linewidth]{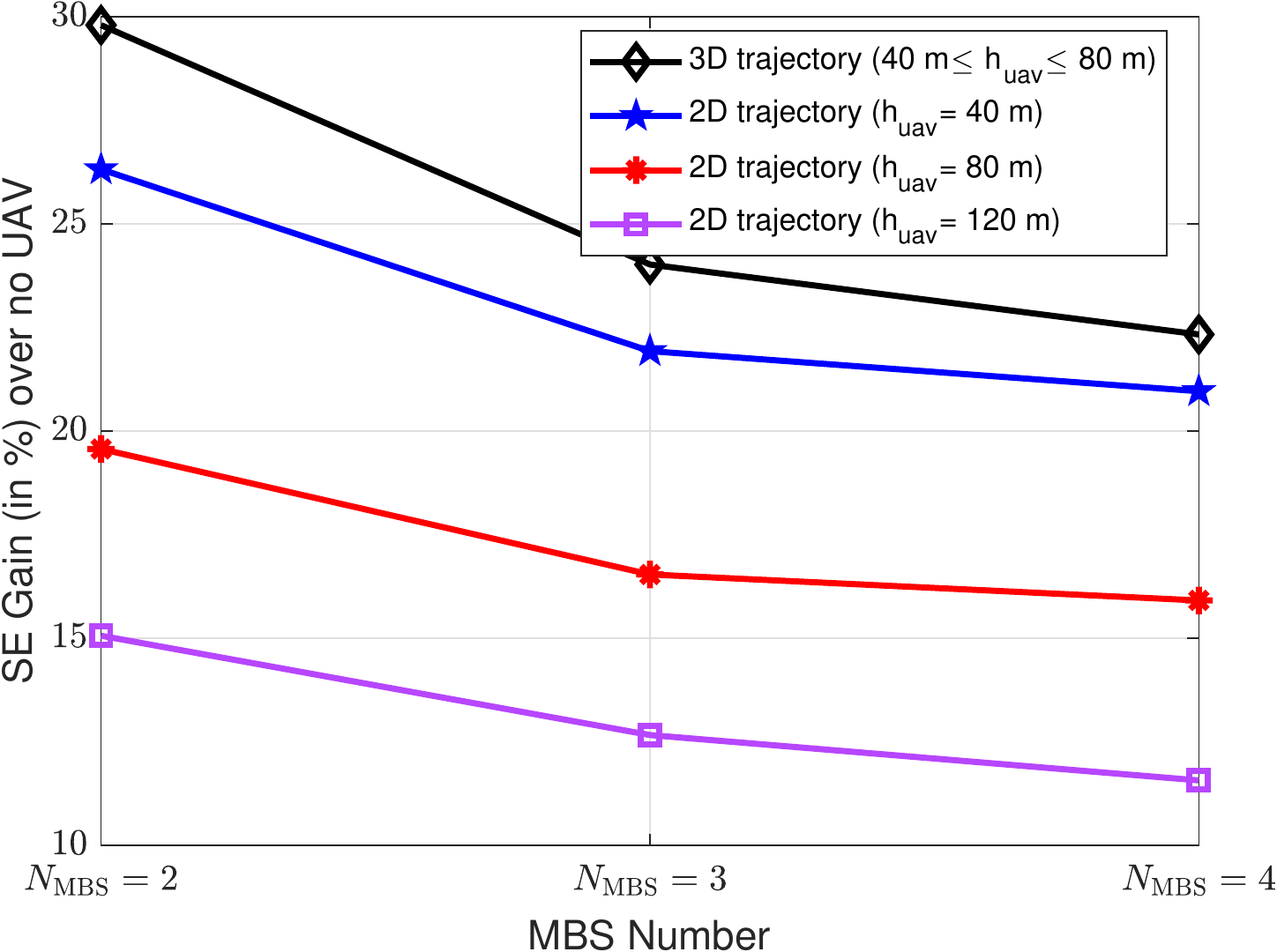}
\caption{SE gain (in \%) comparison between 3D trajectory and 2D trajectories with 40 m, 80 m, and 120 m UAV height for $T = 240 $~s.}
\label{SE_gain}
\end{figure}
In order to demonstrate the advantage of changing altitudes for getting better antenna gain, we also compute the 2D optimal paths for three different fixed heights (40 m, 80 m and 120 m). Then we calculate the per UE SE and plot the gains over no UAV case to compare them with that of 3D trajectory in Fig.~\ref{SE_gain}. We can conclude that all of the trajectories provide significant gain with respect to no UAV scenarios. The gains overall tend to decrease with the increasing number of MBSs, since a higher number of MBSs provide the UEs better chance of getting high antenna gain (around 10 to 15 dB). As a result, the UAV can not take over a UE easily from its serving MBS which translates into lower SE gain over no UAV case for a higher number of MBS.

The 3D trajectory in Fig. \ref{SE_gain} outperforms the other 2D trajectories in terms of SE gain. This is due to the the capability of changing heights, which provides the UAV higher antenna gain from the sidelobes lying between 270\degree~ and 0\degree~, as depicted in Fig.~\ref{antenaa_radiation}(a). The other 2D counterparts have less flexibility to increase the antenna gain in the backhaul link and hence, provide lower SE gain. Among the 2D trajectories, the one with 40 m height provides higher SE gain than the trajectories associated with 80 m and 120 m. This is due to the fact that a lower height can provide larger elevation angle with respect to the $z$ axis and thus, can obtain higher gain by the two sidelobes lying between 330\degree~and 0\degree. The 2D trajectories pertinent to higher altitudes have less chance to get gain through the sidelobes closer to the mainlobe. This phenomenon is also reflected in Fig.~\ref{antenaa_radiation}(b).  
\begin{figure}
\centering
\includegraphics[width=.85\linewidth]{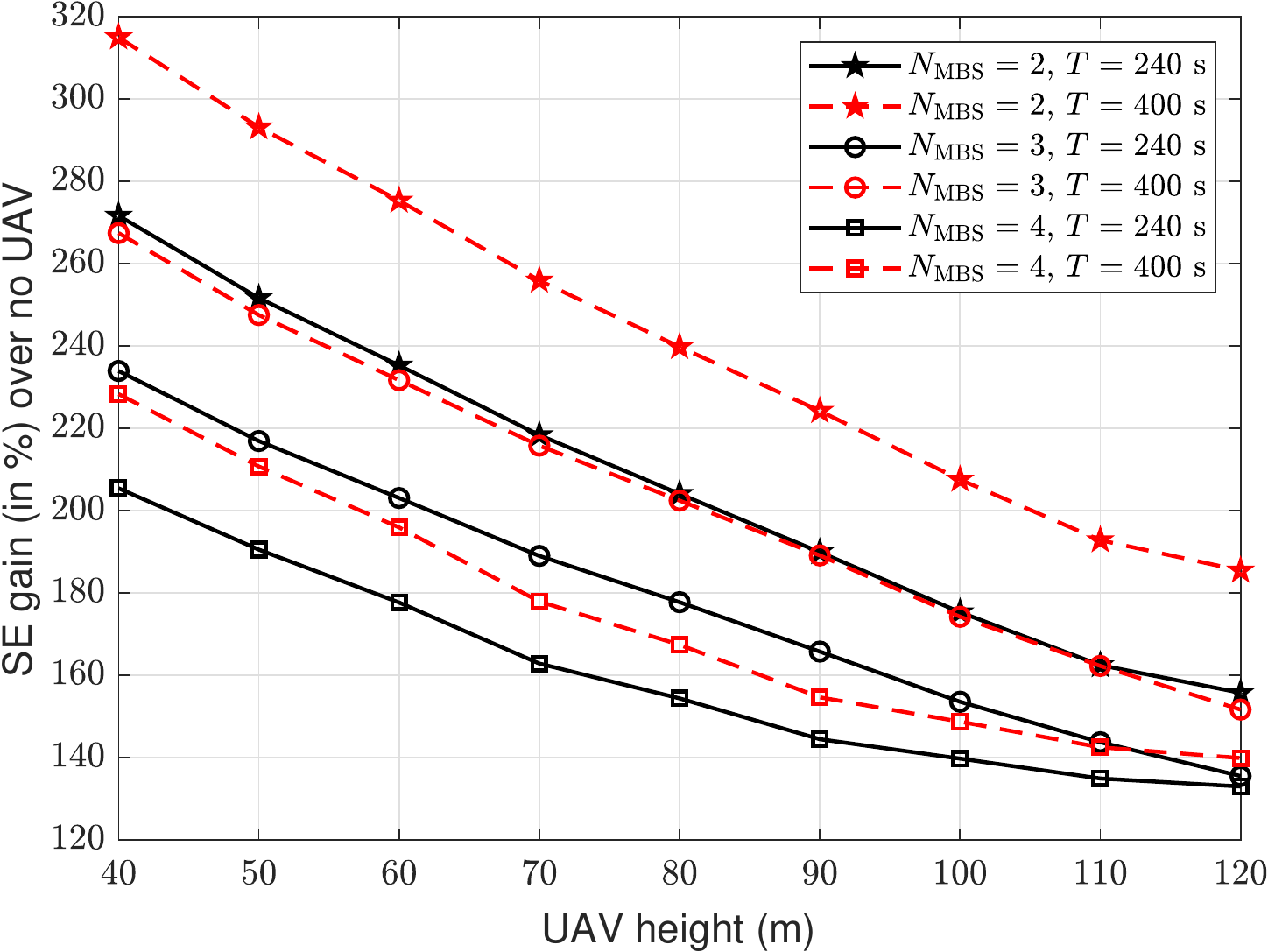}\\
\caption{Comparison of SE gain (in \%) over no UAV vs UAV  height (m) for 2D trajectory optimization ($T = 240 $~s and $T = 400 $~s).}
\label{optimal_height_6}
\end{figure}

For 2D trajectory optimization, we also study the relation between UAV heights and SE gains over no UAV scenario in Fig.~\ref{optimal_height_6} which highlights the SE gain over no UAV with respect to different UAV heights for different values of $N_{\text{MBS}}$ and $T$. As expected, the SE gains pertinent to $N_{\text{MBS}} = 2$ and $N_{\text{MBS}}=3$ tend to plummet with increasing UAV height due to low antenna gains in the backhaul connection. Since a longer $T$ allows the UAV to spend more time on the optimal points, $T=400 $ s provides higher SE gains than $T=240 $ s. 

\subsection{Outage Probability Comparison}
Next, we study the outage probability of the optimal trajectories for $T=240 $ s and $\lambda_{\text{ue}}=50$, where outage probability of a UE $k$ is defined as,
\begin{equation}
    P_{\text{out}}=\mathbb{P}\{R_k(i)< t_c\}\; \forall\; i\in{0,1,..,N},
\end{equation}
where $t_c$ is the threshold for minimum quality of service (QoS) requirement. This threshold is considered as 0.05 bps/Hz in our study. We determine the number of UEs with SE less than the threshold in each time step of the trajectory and then calculate the mean over the mission duration. After that, we divide the number of UEs with lower SE than $t_c$ over the mission duration by the total number of UEs to obtain the outage probability. We then repeat the similar steps for each random network associated with different values of $\lambda_{\text{mbs}}$. We also calculate the outage probability of each same network without considering any UAV for comparison purposes. 

\begin{figure}[t]
\centering
\includegraphics[width=0.85\linewidth]{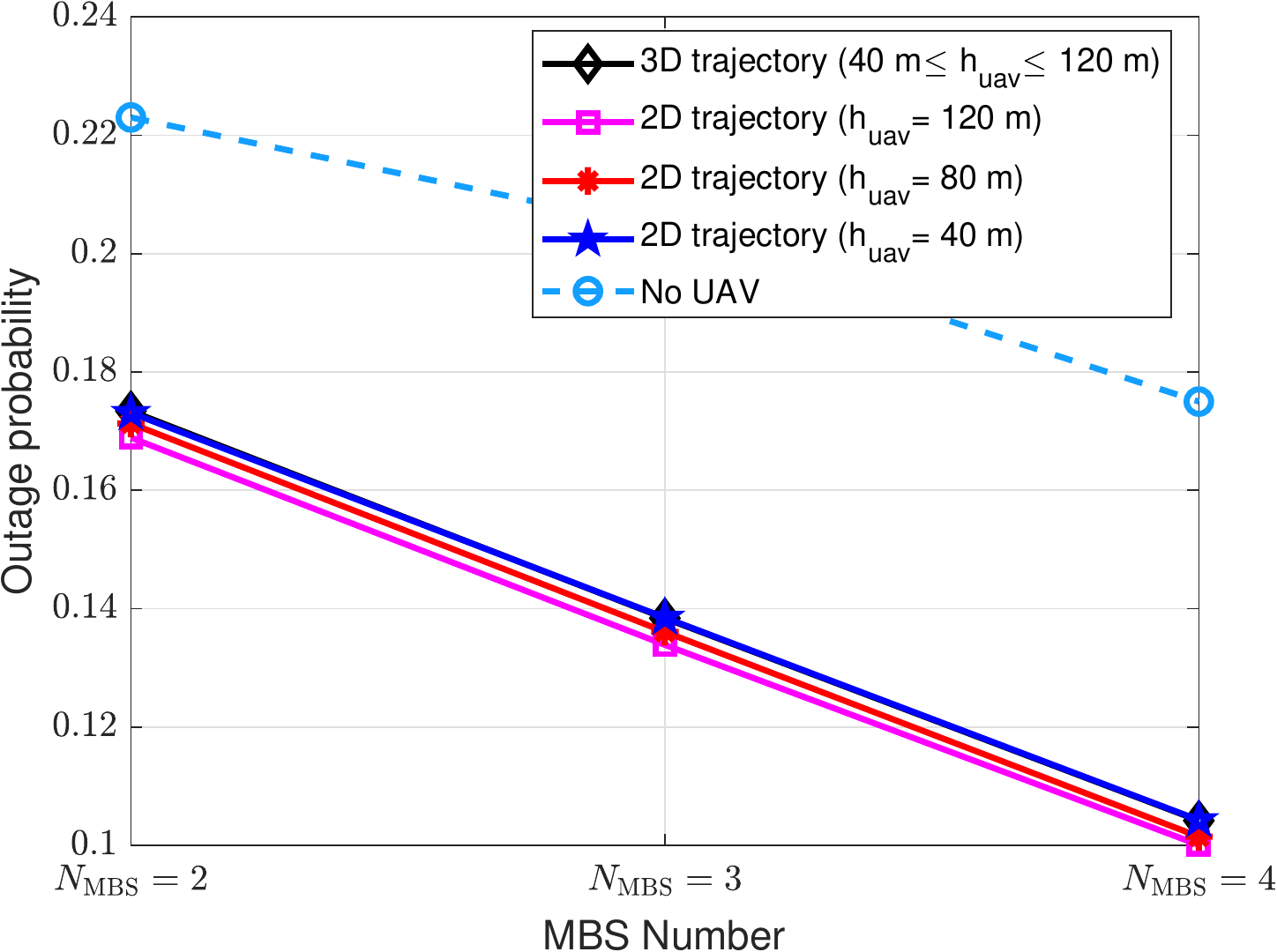}
\caption{{Outage probability comparison between 3D trajectory and 2D trajectories with 40 m, 80 m, and 120m UAV height for $T= 240$s.  }}
\label{out_gain}
\end{figure}

The relevant results are shown in Fig.~\ref{out_gain} which highlights that the 3D trajectory provides  similar coverage performance as  the other 2D trajectories. Even though the 3D trajectory provides higher SE gain, it also creates more interference to ground UEs and hence provides slightly higher outage probability than its 80 m and 120 m counterparts. The 2D trajectory associated with 40 m height coincides with the 3D trajectory since it also generates higher interference power due to the higher exposure probability to the sidelobes with a relatively higher gain and its low altitude. Overall, the presence of UAV improves the coverage of the networks. Outage probabilities decrease with the increasing number of MBSs due to better coverage and SIR, not to mention the higher chance of getting better antenna gains from the MBSs. Apart from this, the UAV can also get higher gain in the backhaul link due to larger number of MBSs and hence, the end-to-end SIR of the relay links improve. The mixture of all these effects results in better coverage throughout the network.

\subsection{5pSE Comparison}

\begin{figure}[t!]
\centering
\includegraphics[width=.85\linewidth]{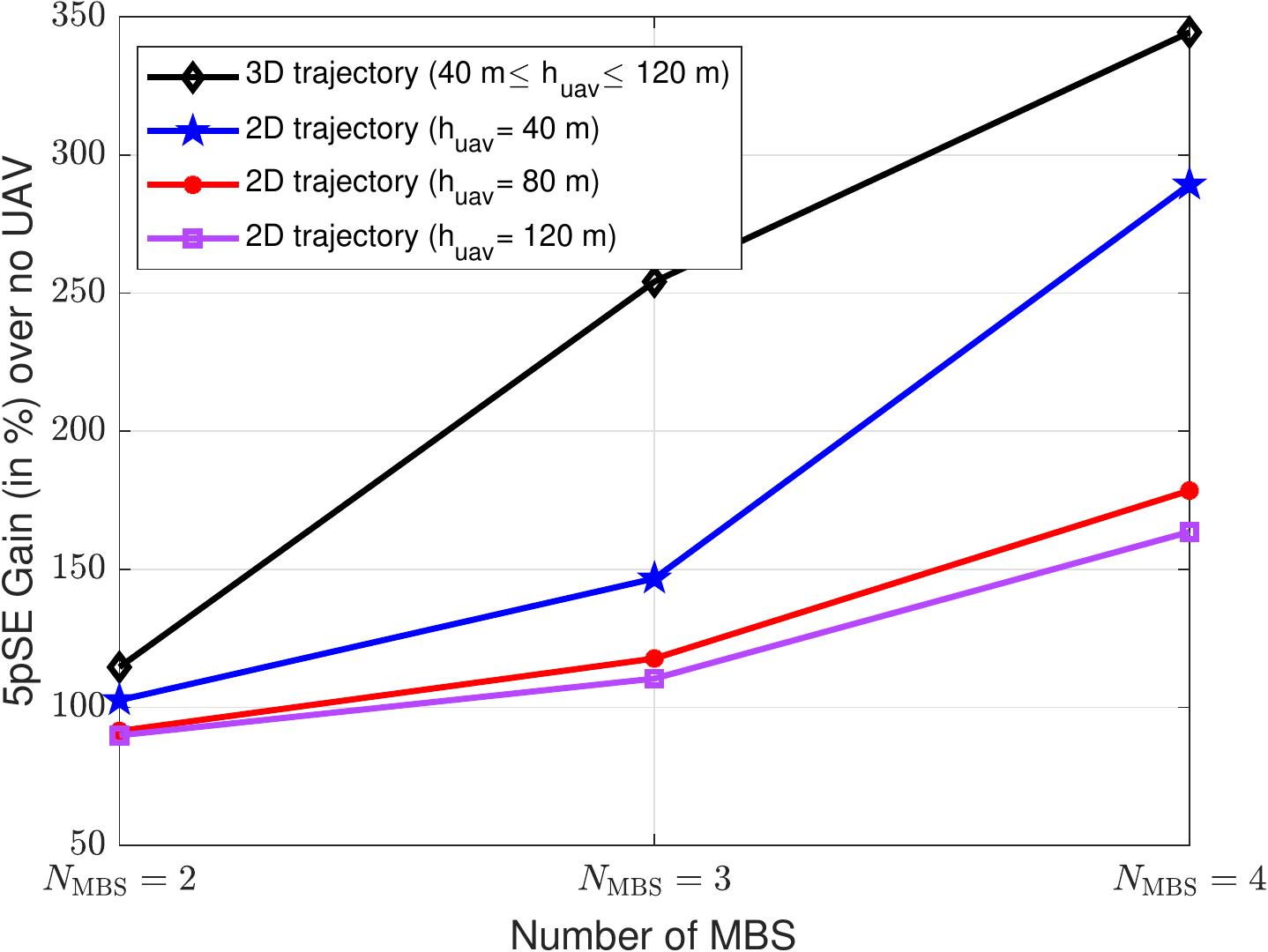}\\
\caption{5pSE comparison between 3D trajectory and 2D trajectories with 40 m, 80 m, and 120 m UAV height for $T= 240$s.}
\label{5pSE}
\end{figure}

Maintaining a minimum QoS level for all of the UEs in a cell is an important deployment aspect for existing cellular networks. The fifth percentile SE (5pSE) takes the worst fifth percentile UE SE of the network into account. In this study, we calculate the time-averaged 5pSE associated with the optimal 3D trajectories of all random networks with $\lambda_{\text{ue}}=20$ and take the mean. We also compute the 5pSE of the same random networks considering 2D trajectory optimization. The 5pSE associated with no UAV scenario is also computed to calculate the gain of optimal paths over the no UAV case. The results are depicted in Fig.~\ref{5pSE}, from which we can conclude that the UAV presence can indeed increase 5pSE to a significant extent. The optimal 3D trajectory can increase the 5pSE of the network by more than 100\%. The 5pSE gains also increase with the number of MBSs due to the fact that more MBSs can provide higher antenna gain in the backhaul link and thus help the UAV to provide coverage assistance to the most deprived UEs. Here, the 3D trajectory outperforms the 2D trajectories due to the flexibility of obtaining higher gains in the backhaul link. As expected, 2D trajectory at 40 m UAV height provides the higher 5pSE gain when compared to the other two. 
 
\begin{table}[!t]
\centering
\renewcommand{\arraystretch}{1.2}
\caption {Number of discrete UAV heights versus execution time.}
\scalebox{1.05}
{\begin{tabular}{cc}
\hline
Number of discrete UAV height levels & Execution time (s) \\
\hline
2 & 0.6288  \\
3 & 0.9135  \\
4 & 1.2608 \\
5 & 1.7090 \\ 
6 & 2.2254\\
7 &  2.8697\\
\hline
\end{tabular}}
\label{height}
\end{table}

\subsection{Execution Time Analysis}
\begin{figure}[t]
\centering
\includegraphics[width=0.85\linewidth]{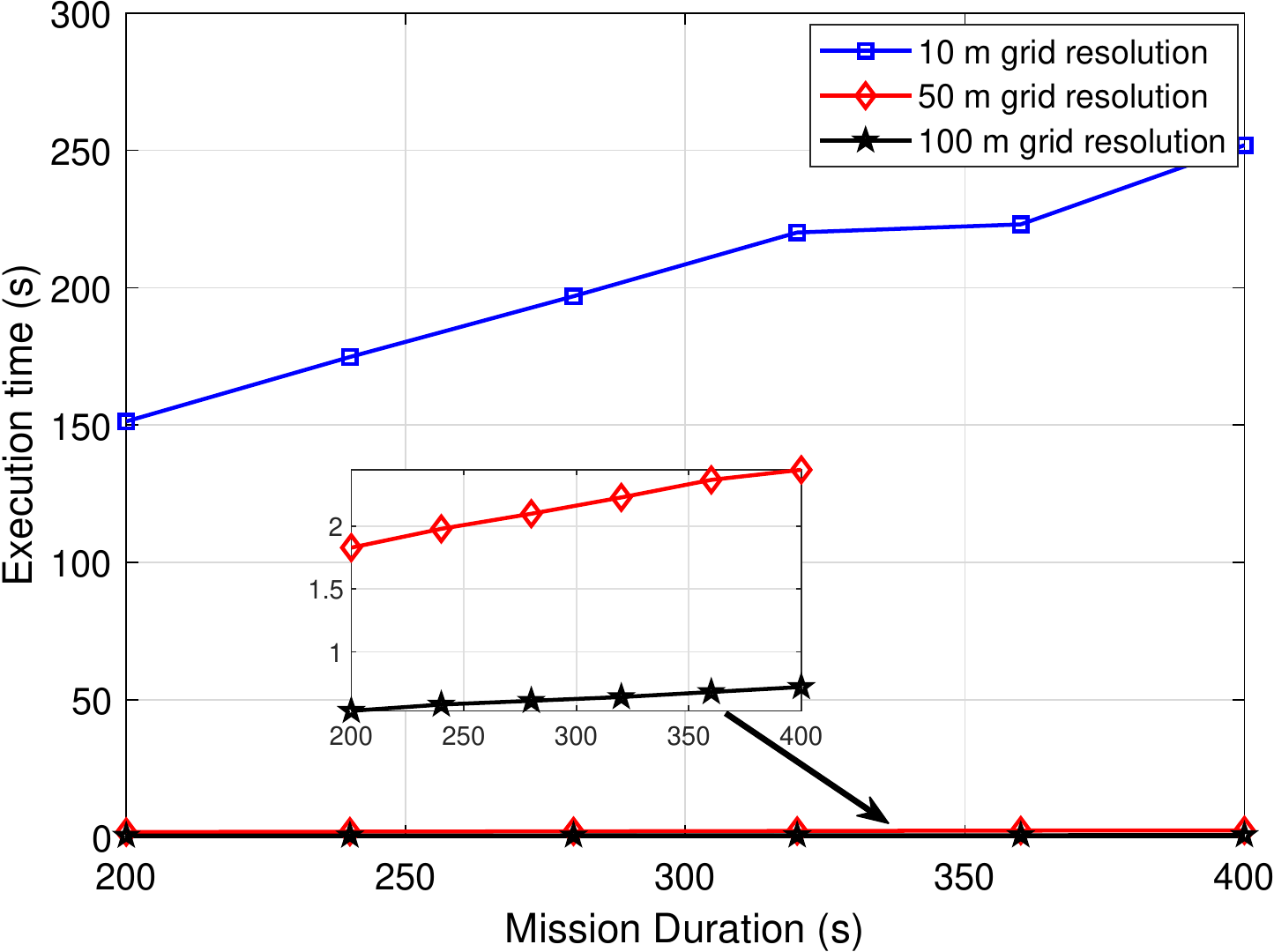}
\caption{Execution time (s) vs mission duration $T$ comparison for different grid resolutions along the $xy$ plane.}
\label{comp}
\end{figure}

In Fig. \ref{comp}, we plot the execution time of the 3D path planning problems with different values of $T$ and grid resolutions along the $xy$ plane. We consider three different grid resolutions: 10 m, 50 m, and 100 m. We also consider 40 m  and 50 m as discrete UAV heights. As expected, the higher grid resolutions translate into higher execution times. Trajectory optimization with 10 m grid resolution needs a significant amount of computation time than its other counterparts. The time taken by 50 m grid resolution increases with a steeper slope than its 100 m counterpart. Note that, in line with our analysis in Section VI.A, the execution times increase almost linearly with $T$. Increasing grid resolution will give a solution very close to the optimal one \cite{Bulut}, though it will come at a cost of very expensive computation. We also calculate the execution time for a different number of discrete UAV heights that are allowed in 3D trajectory optimization. The results in Table \ref{height} show that increasing the number of possible UAV heights increases the number of states, which translates into higher computation time. It is worth noting that for 2D trajectory optimization we need to consider only one discrete UAV height. We use Intel(R) Core(TM) i7-4770 CPU with 3.40 GHz clock speed, 16 GB RAM and x-64 based processor workstation for calculating the optimal paths in Matlab 2018b.
\subsection{Effects of Antenna Downtilt Angle}

\begin{figure}
		\centering
		\subfloat[]{
			\includegraphics[width=.85\linewidth]{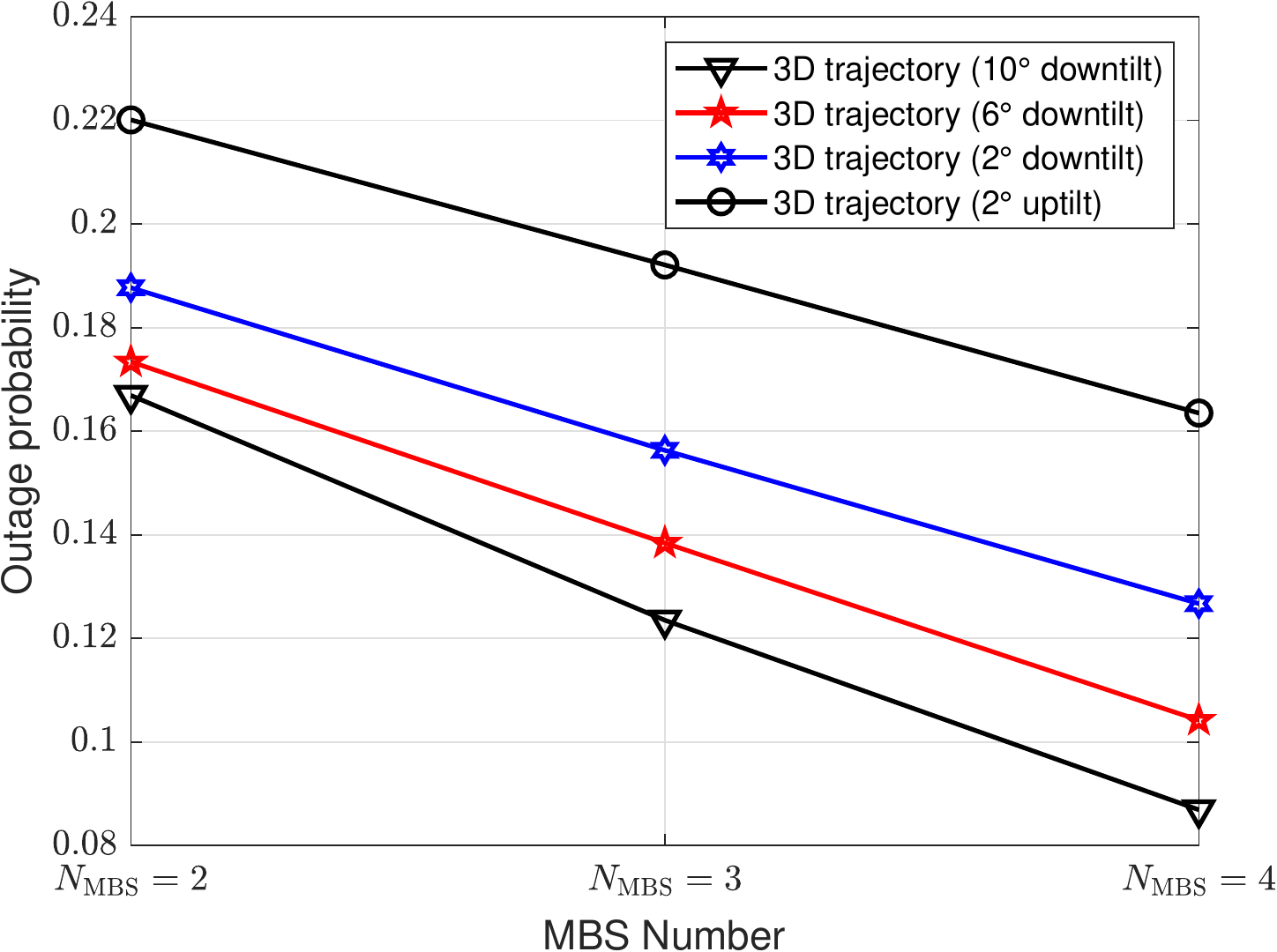}}
           		
		\subfloat[]{
			\includegraphics[width=.85\linewidth]{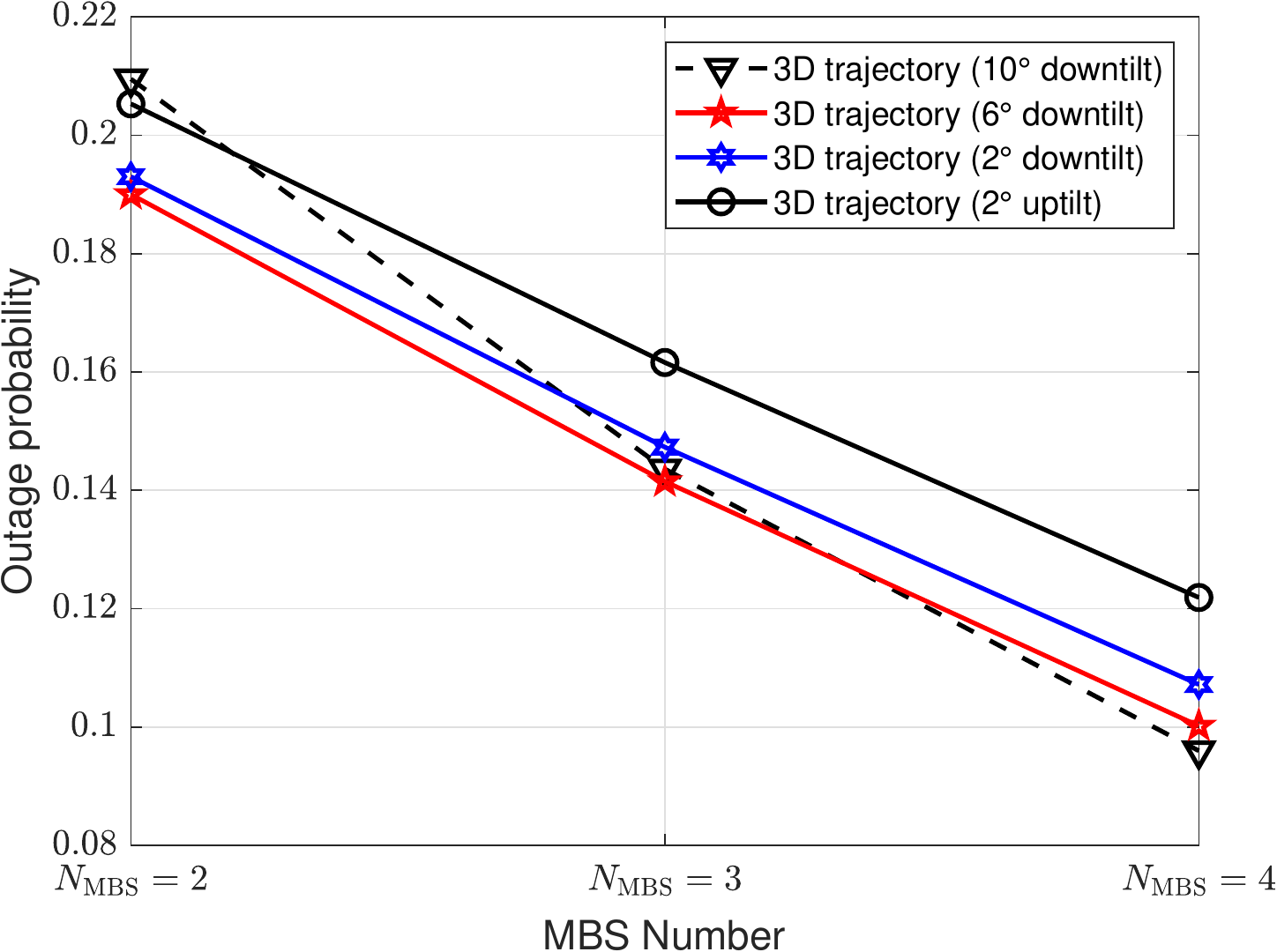}}
			
		\caption {
		{Outage probability comparison between 3D trajectories with different antenna tilting angles for $T= 240$s. (a) Simulation area $= 1.2 \times 1.2~ \text{km}^2$.
		(b) Simulation area $= 2.2 \times 2.2 ~\text{km}^2$.
        }}
		\label{tilt_angle}
\end{figure}

According to \cite{edt}, in an interference-limited scenario, electrical downtilt provides better SINR than mechanical downtilt. Moreover, the system throughput also depends significantly on the downtilt angle. Hence, in this subsection, we study the outage performance for the optimal 3D trajectories for different electrical downtilt angles and show the associated results in Fig.~\ref{tilt_angle}. In Fig.~\ref{tilt_angle}(a), we show the outage probabilities associated with different downtilt angles for different $N_{\text{MBS}}$ for simulation area $= 1.2 \times 1.2~ \text{km}^2$. We can conclude that the outage probabilities decrease almost linearly with increasing $N_{\text{MBS}}$. We can also conclude that lowering the tilting angle decreases the outage probability. This is because lowering the tilt angle decreases interference towards other nearby MBSs which translates into better coverage for the UEs. Directing the antenna mainlobe towards the UAV by changing the tilt angle upwards ($-2^{\circ}$) degrades the outage performance. Even though the UAV will enjoy better connectivity in the MBS-to-UAV link, UEs in the ground will be served by the antenna sidelobes which will translate into higher outage probability since the majority of the UEs are served by the MBSs.

To study the effects of downtilt angles on outage probability more closely, we run simulations for area $= 2.2 \times 2.2~ \text{km}^2$ while keeping the number of MBSs and UEs same. We depict the corresponding results in Fig.~\ref{tilt_angle}(b) from which we observe that for lower MBS density, higher downtilt angle is not always beneficial since MBSs need to cover a larger user area.
~For $N_{\text{MBS}}=2$, $10^{\circ}$ angle provides the lowest coverage performance since it focuses on a smaller portion of the network than its other counterparts. Both $6^{\circ}$ and $2^{\circ}$ can provide high antenna gains towards a larger area and hence, provide lower outage probabilities for $N_{\text{MBS}}=2$. As expected, with increasing $N_{\text{MBS}}$, the probability of getting high antenna gains at the UEs increases, which translates into lower outage probabilities for all downtilt angles. For $N_{\text{MBS}}=4$, $10^{\circ}$ angle provides the best coverage performance since it needs to cover less area and associate fewer UEs residing nearby.

\section{Concluding Remarks}
\label{sec:conclusion}
In this paper, we study the effects of backhaul constraints and 3D antenna radiation pattern on optimal path planning in interference prevalent downlink cellular networks. We formulated the optimal 3D trajectory calculation problem and solved it using DP. Through extensive simulations we show that the presence of UAVs will provide significant capacity and 5pSE gain over no UAV scenario. 3D trajectory can exploit more degrees of freedom than its 2D counterparts to provide better network performance. Due to backhaul constraint, the UAV has less freedom to move around the network and thus 3D trajectory can help to increase SE gain and 5pSE rate by exploiting more freedom to achieve better antenna gain in the backhaul link than 2D trajectory, at the cost of higher computational complexity.

\bibliographystyle{myIEEEtran} 
\bibliography{ref.bib}

\end{document}